\documentclass[12pt]{article}

\usepackage{amsfonts}
\usepackage[left=1in, right=0.7in, top=1in, bottom=1in, nohead,
  a4paper]{geometry}

\def\asl3{\widehat{\mathfrak{sl}_3(\mathbb C)}}
\def\bbC{\mathbb C}
\def\bbZ{\mathbb Z}
\def\iu{\mathrm i\,}
\def\gg{\mathfrak g}
\def\rme{\mathrm e}
\mathchardef\Delta="101
\mathchardef\Gamma="100
\mathchardef\Phi="108
\mathchardef\Sigma="106
\mathchardef\Theta="102

\makeatletter
\def\eqnarray{%
   \stepcounter{equation}%
   \def\@currentlabel{\p@equation\theequation}%
   \global\@eqnswtrue
   \m@th
   \global\@eqcnt\z@
   \tabskip\@centering
   \let\\\@eqncr
   $$\everycr{}\halign to\displaywidth\bgroup
       \hskip\@centering$\displaystyle\tabskip\z@skip{##}$\@eqnsel
      &\global\@eqcnt\@ne\hfil$\displaystyle{\hbox{}##\hbox{}}$\hfil
      &\global\@eqcnt\tw@ $\displaystyle{##}$\hfil\tabskip\@centering
      &\global\@eqcnt\thr@@ \hb@xt@\z@\bgroup\hss##\egroup
         \tabskip\z@skip
      \cr
}
\makeatother
\catcode`\@=11
\def\marginnote#1{}
\newcount\hour
\newcount\minute
\newtoks\amorpm
\hour=\time\divide\hour by60
\minute=\time{\multiply\hour by60 \global\advance\minute
by-\hour}\edef\standardtime{{\ifnum\hour<12
\global\amorpm={am}%
        \else\global\amorpm={pm}\advance\hour by-12 \fi
        \ifnum\hour=0 \hour=12 \fi
        \number\hour:\ifnum\minute<10
0\fi\number\minute\the\amorpm}}
\edef\militarytime{\number\hour:\ifnum\minute<10
0\fi\number\minute}

\def\draftlabel#1{{\@bsphack\if@filesw {\let\thepage\relax
   \xdef\@gtempa{\write\@auxout{\string
      \newlabel{#1}{{\@currentlabel}{\thepage}}}}}\@gtempa
   \if@nobreak \ifvmode\nobreak\fi\fi\fi\@esphack}
        \gdef\@eqnlabel{#1}}
\def\@eqnlabel{}
\def\@vacuum{}
\def\draftmarginnote#1{\marginpar{\raggedright\scriptsize\tt#1}}
\def\draft{\oddsidemargin -.5truein
        \def\@oddfoot{\sl preliminary draft \hfil
        \rm\thepage\hfil\sl\today\quad\militarytime}
        \let\@evenfoot\@oddfoot \overfullrule 3pt
        \let\label=\draftlabel
        \let\marginnote=\draftmarginnote

\def\@eqnnum{(\theequation)\rlap{\kern\marginparsep\tt\@eqnlabel}%
\global\let\@eqnlabel\@vacuum}  }

\def\numberbysection{\@addtoreset{equation}{section}
        \def\theequation{\thesection.\arabic{equation}}}

\def\underline#1{\relax\ifmmode\@@underline#1\else
 $\@@underline{\hbox{#1}}$\relax\fi}

\catcode`@=12

\numberbysection

\def\sect#1{\section{#1}}

\def\rf#1{(\ref{#1})}
\def\lab#1{\label{#1}}
\def\nonu{\nonumber}
\def\br{\begin{eqnarray}}
\def\er{\end{eqnarray}}
\def\be{\begin{equation}}
\def\ee{\end{equation}}

\def\lb{\lbrack}
\def\rb{\rbrack}

\def\({\left(}
\def\){\right)}

\newcommand{\ct}[1]{\cite{#1}}
\newcommand{\bi}[1]{\bibitem{#1}}

\relax

\newcommand\sbr[2]{[\,{#1},\,{#2}\,]}

\newcommand\bra[1]{\langle \, {#1}\, |}
\newcommand\ket[1]{| \, {#1} \, \rangle}
\newcommand\braopket[3]{\langle \, #1 \, | \, #2 \, | \, #3 \, \rangle}

\def\a{\alpha}

\def\b{\beta}

\def\d{\delta}

\def\eps{\epsilon}

\def\g{\gamma}
\def\G{\Gamma}

\def\h{{1\over 2}}
\def\l{\lambda}

\def\nut{{\widetilde   \nu}}

\def\pa{\partial}
\def\pr{\prime}

\def\psit{{\widetilde \psi}}

\def\ra{\rightarrow}

\def\tp0{\Theta_{+}^{(0)}}
\def\tm0{\Theta_{-}^{(0)}}

\def\ut{{\widetilde  U}}
\def\vp{\varphi}

\def\cgh{\mathfrak g}

\def\tone{\tau_0}
\def\ttwo{\tau_1}
\def\tthree{\tau_2}
\def\tfour{{\tilde \tau}_R^{\a_1}}
\def\tfive{{\tilde \tau}_R^{\a_2}}
\def\tsix{\tau_R^{\a_3}}
\def\tseven{\tau_L^{\a_1}}
\def\teight{\tau_L^{\a_2}}
\def\tnine{{\tilde \tau}_L^{\a_3}}
\def\tten{{\tilde \tau}_{L,(2)}^{\a_1}}
\def\televen{{\tilde \tau}_{L,(0)}^{\a_1}}
\def\ttwelve{{\tilde \tau}_{L,(1)}^{\a_2}}
\def\tthirteen{{\tilde \tau}_{L,(0)}^{\a_2}}
\def\tfourteen{\tau_{L,(1)}^{\a_3}}
\def\tfifteen{\tau_{L,(2)}^{\a_3}}
\def\tsixteen{\tau_{R,(0)}^{\a_1}}
\def\tseventeen{\tau_{R,(2)}^{\a_1}}
\def\teighteen{\tau_{R,(0)}^{\a_2}}
\def\tnineteen{\tau_{R,(1)}^{\a_2}}
\def\ttwenty{{\tilde \tau}_{R,(1)}^{\a_3}}
\def\ttwentyone{{\tilde \tau}_{R,(2)}^{\a_3}}

\def\f#1#2#3 {f^{#1#2}_{#3}}

\def\win1{{\sf w_{1+\infty}}}

\def\Win1{{\sf W_{1+\infty}}}

\def\rlx{\relax\leavevmode}
\def\inbar{\vrule height1.5ex width.4pt depth0pt}
\def\IZ{\rlx\hbox{\sf Z\kern-.4em Z}}
\def\IR{\rlx\hbox{\rm I\kern-.18em R}}
\def\IC{\rlx\hbox{\,$\inbar\kern-.3em{\rm C}$}}
\def\IN{\rlx\hbox{\rm I\kern-.18em N}}
\def\IO{\rlx\hbox{\,$\inbar\kern-.3em{\rm O}$}}
\def\IP{\rlx\hbox{\rm I\kern-.18em P}}
\def\IQ{\rlx\hbox{\,$\inbar\kern-.3em{\rm Q}$}}
\def\IF{\rlx\hbox{\rm I\kern-.18em F}}
\def\IG{\rlx\hbox{\,$\inbar\kern-.3em{\rm G}$}}
\def\IH{\rlx\hbox{\rm I\kern-.18em H}}
\def\II{\rlx\hbox{\rm I\kern-.18em I}}
\def\IK{\rlx\hbox{\rm I\kern-.18em K}}
\def\IL{\rlx\hbox{\rm I\kern-.18em L}}
\def\one{\hbox{{1}\kern-.25em\hbox{l}}}
\def\0#1{\relax\ifmmode\mathaccent"7017{#1}%
B        \else\accent23#1\relax\fi}

\def\NPB#1#2#3{{\sl Nucl. Phys.} {\bf B#1} (#2) #3}

\def\CMP#1#2#3{{\sl Commun. Math. Phys.} {\bf #1} (#2) #3}

\def\PLB#1#2#3{{\sl Phys. Lett.} {\bf #1B} (#2) #3}
\def\JMP#1#2#3{{\sl J. Math. Phys.} {\bf #1} (#2) #3}

\def\FAaIA#1#2#3{{\sl Funct. Anal. Appl.} {\bf #1} (#2) #3}

\def\IJMPA#1#2#3{{\sl Int. J. Mod. Phys.} {\bf A#1} (#2) #3}

\def\PHSD#1#2#3{{\sl Physica} {\bf D#1} (#2) #3}
\def\PJA#1#2#3{{\sl Proc. Japan. Acad.} {\bf #1A} (#2) #3}

\begin{document}
\begin{titlepage}
\vspace*{-1cm}

\noindent
May, 2001 \hfill{IFT-P.040/01}\\
\hfill{hep-th/0105078}

\vskip 3cm

\vspace{.2in}
\begin{center}
{\large\bf\mathversion{bold}Confinement and soliton solutions in the
$SL(3)$ Toda model coupled to matter fields}
\end{center}

\vspace{.5in}

\begin{center}
A. G. Bueno~$^1$, L. A. Ferreira~$^1$ and A. V. Razumov~$^2$

\vspace{.5 cm}
\small

\par \vskip .2in \noindent
$^{(1)}$~Instituto de F\'\i sica Te\'orica - IFT/UNESP\\
Rua Pamplona 145\\
01405-900  S\~ao Paulo-SP, Brazil\\

\par \vskip .2in \noindent
$^{(2)}$~Institute for Hight Energy Physics\\
142280 Protvino, Moscow Region, Russia

\normalsize
\end{center}

\vspace{.5 in}

\begin{abstract}

We consider an integrable conformally invariant two dimensional model
associated to the affine Kac-Moody algebra $\asl3$. It possesses four
scalar fields and six Dirac spinors. The theory does not possesses a local 
Lagrangian since the spinor equations of motion present interaction terms
which are bilinear in the spinors. There exists a submodel presenting an
equivalence between a $U(1)$ vector current and a topological
current, which leads to a confinement of the spinors inside the solitons.
We calculate the one-soliton and two-soliton solutions using a procedure
which is a hybrid of the dressing and Hirota methods. The soliton
masses and time delays due to the soliton interactions are also calculated.
We give a computer program to calculate the soliton
solutions.

\end{abstract}
\end{titlepage}

\sect{Introduction}

The non-perturbative aspects of Lorentz invariant field theories are
related in one or another way to classical soliton
solutions. The best known example is provided by the strong coupling sector
of non-abelian gauge theories. It is believed that the fundamental
particles, namely gluons and quarks, get confined and form bound states
which should correspond to the known spectrum of hadrons. The most popular
mechanism to explain such confinement is based on a condensation of
magnetic monopoles producing a vacuum which is a magnetic superconductor.
The quarks and gluons get then confined by the dual Meissner effect.
The monopoles responsible for that mechanism are not  excitations of the
fundamental fields appearing in the Lagrangian. They are classical soliton
solutions carrying a topological charge. There are many ideas based on
Montonen-Olive duality conjecture \ct{montonen} where the monopoles are
interpreted as the fundamental particles of a dual Lagrangian describing
the strong coupling sector of gauge theories.

The development of exact methods for studying solitons and non-perturbative
aspects of field theories is important for many reasons. Besides its
relevance for gauge theories, it finds applications in many areas of
condensed matter physics, non-linear phenomena and fluid dynamics. In many
cases, the symmetries involved allow the reduction of the number of
dimensions, and integrable non-linear models in low dimensions play an
important role. The soliton theory in  two dimensions has
reached a high degree of development. It is known  how to construct
practically all the soliton solutions of two dimensional models
using some very basic methods. Such theories have a zero curvature
representation with the potentials taking values on an affine Kac-Moody
algebra. In addition, it should present some ``vacuum'' solutions such that
the zero curvature potentials belong to an abelian subalgebra when
evaluated on them. The soliton solutions are then constructed via the
dressing method \ct{dressing} where the dressing transformation are
generated by the operators which diagonalize the adjoint action of the
``oscillators'' of that abelian subalgebra \ct{tau}. Some of the structures
of two-dimensional integrable models can in fact be generalized to theories
in higher dimensions \ct{afsg}.

In this paper we consider a two-dimensional field theory involving four
scalar fields and six Dirac spinors. The interactions among the fields
present some peculiar properties. There are the usual coupling of bilinears
in the spinors to exponentials of the scalars. However, some of the
equations of motion for the spinors present terms which are bilinear in the
spinors themselves. That fact makes it difficult to believe that one can
find a local Lagrangian for the theory. In spite of that fact, the model
presents a lot of symmetries. It is conformally invariant, possesses local
gauge symetries as well as vector and axial conserved currents bilinear in
the spinors. In addition, it presents three species of one-soliton
solutions carrying non-trivial topological charges. One of the most
striking properties of the model is that it presents a confinement
mechanism of the spinors inside the solitons. For a special submodel there
exists an equivalence between a $U(1)$ vector conserved current,
bilinear in the spinors, and a topological currents depending only on the
first derivative of some scalars. The equivalence is such that the charge
density associated to that $U(1)$ current can only exist on those regions
where the space derivative of the scalars are non-vanishing. If one then
consider excitations of the fields around a given soliton solution, one
observes the charge density can exists only inside the solitons, since the
scalars tend to constants outside them. Therefore outside the solitons,
the spinors can live in ``white'' states carrying no $U(1)$ charge. That
resembles very much what one expexts to happen with the confinement
mechanism of QCD, even though our model presents a bag model like
confinement instead of the dual Meissner effect.

The model possesses a zero curvature representation based on the
$\widehat{\mathfrak{sl}_3(\mathbb C)}$ affine kac-Moody algebra.
It constitutes a particular example of the so-called  affine Toda
models coupled to matter fields which have been introduced in paper
\ct{matter}. The corresponding model associated to
$\widehat{\mathfrak{sl}_2(\mathbb C)}$ has been studied in paper
\ct{su2matter} where it was shown, using bosonization methods, that the
equivalence between the currents holds true at the quantum level and so the
confinement mechanism does take place in the quantum theory.

The objective of the present paper is to calculate exactly the three
species of one-soliton solutions as well as the corrensponding six
two-soliton solutions describing the scattering between any two of them. We
also evaluate the time-delays for the interactions among the solitons. The
character of the interactions, i.e. the relative strength and if it is
attractive or repulsive, is determined by the scalar product of the roots
associated to the solution. The solutions are constructed using a mixture
of the dressing \ct{dressing} and Hirota \ct{hirota} methods. The dressing
method in association to the vertex operator representations of the
Kac-Moody algebra is quite good in finding the spectrum of solitons, to
construct the so-called tau-functions \ct{tau}, and to provide an ansatz
for the solution. The Hirota method, which does not provide us any hint of
the correct tau-functions, is then used to actually  evaluate
the explicit expression of the soliton solutions.   We have implemented an
algorithm in {\sc Mathematica} to execute the procedures of the Hirota
method. In this way we  escape from two difficulties of the
methods. The Hirota method needs the tau-functions and an ansatz for them,
but it does not tell us how to construct those. That is solved through the
dressing method, which in its turn needs the evaluation of matrix elements
in the vertex operator representation.  We avoid that calculation through
the Hirota expansion method.

The paper is organized as follows. In Section \ref{sec:description} we
describe in detail the model, its symmetries and the zero curvature
representation. In Section \ref{sec:dressing} we discuss the dressing and
Hirota methods, construct the tau-functions and the corresponding Hirota's
equations. In Section \ref{sec:topologicalcharge} we discuss how to use the
infinetely degenerate vacua to introduce non-trivial topological
charges. Although we work with a $\widehat{\mathfrak{sl}_3(\mathbb C)}$
Kac-Moody algebra, the topological charges take values in the weight
lattices of three  $\widehat{\mathfrak{sl}_2(\mathbb C)}$ subalgebras. In
Section \ref{sec:masses} we show how to use the breaking of the conformal
symmetry to  evaluate the masses of the solitons just from
their asymptotic behaviour. In Section \ref{sec:timedelays} we show how to
evaluate the time delays associated to the interactions of the solitons in
a two-soliton solution. In Sections \ref{sec:onesol} and \ref{sec:twosol}
the one-soliton and two-soliton solution respectively are calculated.  The
appendix \ref{sec:app} contains the results about Kac-Mooody needed for the
calulations performed in the paper, and in appendix \ref{sec:app2} we
present the algorithm for {\sc Mathematica} used to evaluate the soliton
solutions.

As a matter of conventions, we use for the space-time coordinates in the
two-dimensional Minkowski space the notations $x^0 = t$ and $x^1 = x$. The
light front coordiantes are defined as $x^{\pm}= x^0 \pm x^1$. Hence, we
have $\pa_{\pm} = \h \( \pa_0 \pm \pa_1\)$, and $\pa^2 = \pa_0^2 - \pa_1^2
= 4 \pa_{+}\pa_{-}$. For the $\gamma$-matrices we use the following
representation:
\[
\gamma_0 = -\iu \(
\begin{array}{rr}
0 &-1 \\
1 & 0
\end{array}\), \qquad
\gamma_1 = -\iu \(\begin{array}{rr}
0 & 1 \\
1 & 0
\end{array}\), \qquad
\gamma_5 = \gamma_0\gamma_1 = \(\begin{array}{rr}
1 & 0 \\
0 & -1
\end{array}\).
\]

\newpage

\sect{Description of the model}
\label{sec:description}

The affine Toda models coupled to matter fields have been introduced in
paper \ct{matter}. In the present paper we consider the model
associated to the affine Kac--Mody algebra $\gg = \asl3$, which equations
of motion are\footnote{The spinors have been rescaled by a factor
$\h\sqrt{m_i/\mathrm i}$ with respect to those in \ct{matter}, and the
scalars
$\vp_a$  by a factor $\mathrm i$.}
\br
\pa^2 \vp_a &=&  m_a^2 \; {\bar \psi}^{a}  \,
 V_a \, \left[ \frac{\( 1+\gamma_5\)}{2}   -
e^{3\,   \eta} \frac{\( 1-\gamma_5\)}{2} \) \psi^{a} \nonu\\
&&+  m_3^2 \; {\bar \psi}^{3}  \, V_3 \,
 \( \frac{\( 1+\gamma_5\)}{2}   -
e^{3\,   \eta} \frac{\( 1-\gamma_5\)}{2} \right] \psi^{3},
\qquad a=1,2, \lab{eqmot1}\\
\pa^2 {\widetilde   \nu}   &=& \rme^{3\,  \eta} \left[ \iu \sum_{i=1}^3
m_i^2 \,
{\bar \psi}^{i} \, V_i \,
\frac{\( 1-\gamma_5\)}{2}    \psi^{i}
-  \frac{1}{3}\, \sum_{i=1}^3 m_i^2   \right],   \\
\pa^2 \eta &=& 0, \\
\iu \gamma^{\mu}\pa_{\mu}  \psi^{i}& =&
  m_{i}  \, V_i \,
\,  \left[ \frac{\(  1+\gamma_5\)}{2}
+   e^{3\, \eta}  \frac{\(  1-\gamma_5\)}{2}\right]
\psi^{i} + U^{i}, \qquad i=1,2,3, \\
\iu \gamma^{\mu}\pa_{\mu}  \psit^{i}& =&
 m_{i} \, V_i^{-1} \,
\left[ e^{3\, \eta}  \frac{\( 1+\gamma_5\)}{2}
+ \frac{\( 1-\gamma_5\)}{2} \right] \psit^i
+ {\widetilde  U}^{i}, \qquad i=1,2,3,
\lab{eqmot2}
\er
where $m_3=m_1+m_2$, $\psi^i$ and $\psit^i$, $i=1,2,3$, are independent
Dirac spinors, and ${\bar \psi}^{i}= ( \psit^i )^T \gamma_0$. Notice that
we can not  take
$\psit^i$ to be the complex conjugate of ${\psi^i}^*$, since that is
incompatible with the equations of motion.  In addition, we have
\be
V_1 = \rme^{\iu \gamma_5 \( 2\vp_1 - \vp_2   \)} \,
e^{\gamma_5  \eta }, \qquad
V_2 = \rme^{\iu \gamma_5 \( 2\vp_2 - \vp_1    \)} \,
e^{\gamma_5  \eta }, \qquad
V_3 = \rme^{\iu \gamma_5 \( \vp_1 + \vp_2    \)}\,
e^{2 \, \gamma_5 \eta }.
\lab{potentials}
\ee
Denoting the two component spinors as
\br
\psi^i = \(
\begin{array}{c}
\psi^i_R\\
\psi^i_L
\end{array}\), \qquad \qquad
U^i = \(
\begin{array}{c}
U^i_R\\
U^i_L
\end{array}\),
\er
and similarly for $\psit^i$ and ${\widetilde  U}^{i}$, we  see that
\br
U_R^1 = U_R^2 = U_L^3 = 0, \qquad \ut_L^1 = \ut_L^2 = \ut_R^3 = 0,
\lab{zerous}
\er
and
\br
U_L^1 &=& - \frac{m_2 m_3}{m_1}\, \rme^{\eta}\, \left[
\psi_R^3 \, \psit_L^2 \, e^{i\( 2\vp_2-\vp_1\)} +
\psi_L^3 \, \psit_R^2 \, e^{-i\( \vp_1+\vp_2\)} \right], \lab{nonzerou1}\\
U_L^2 &=&  \frac{m_1 m_3}{m_2}\, \rme^{\eta}\, \left[
\psi_R^3 \, \psit_L^1 \, e^{i\( 2\vp_1-\vp_2\)} +
\psi_L^3 \, \psit_R^1 \, e^{-i\( \vp_1+\vp_2\)} \right], \\
U_R^3 &=&  \frac{m_1 m_2}{m_3}\, \rme^{\eta}\, \left[
\psi_R^2 \, \psi_L^1 \, e^{i\( 2\vp_2-\vp_1\)} -
\psi_L^2 \, \psi_R^1 \, e^{i\( 2\vp_1-\vp_2\)} \right], \\
\ut_R^1 &=& - \frac{m_2 m_3}{m_1}\, \rme^{\eta}\, \left[
\psi_R^2 \, \psit_L^3 \, e^{i\( 2\vp_2-\vp_1\)} +
\psi_L^2 \, \psit_R^3 \, e^{-i\( \vp_1+\vp_2\)} \right], \\
\ut_R^2 &=&  \frac{m_1 m_3}{m_2}\, \rme^{\eta}\, \left[
\psi_R^1 \, \psit_L^3 \, e^{i\( 2\vp_1-\vp_2\)} +
\psi_L^1 \, \psit_R^3 \, e^{-i\( \vp_1+\vp_2\)} \right], \\
\ut_L^3 &=&  \frac{m_1 m_2}{m_3}\, \rme^{\eta}\, \left[
\psit_R^1 \, \psit_L^2 \, e^{i\( 2\vp_2-\vp_1\)} -
\psit_L^1 \, \psit_R^2 \, e^{i\( 2\vp_1-\vp_2\)} \right].
\lab{nonzerou2}
\er
The model under consideration admits a representation in
terms of the Lax-Zakharov-Shabat zero curvature condition \ct{matter}
\be
\pa_{+} A_{-} - \pa_{-} A_{+}  + \lb A_{+}\, , \,  A_{-} \rb = 0,
\lab{zc}
\ee
where the potentials are given by
\br
A_{+} =  - B\, F^{+} \, B^{-1} \, , \qquad
A_{-} = - \pa_{-} B  \, B^{-1} + F^{-}.
\lab{gp}
\er
An important ingredient in the definition of the potentials is the use of
the
principal gradation of $\gg$ (see the definition in the appendix
\ref{sec:app}). The four scalar fields of the theory live in the subgroup
$B$
obtained by exponentiating the zero grade subalgebra $\gg_0$ given in
\rf{zgsa},  and we parametrize it as
\be
B = \rme^{\sum_{a=1}^2 \iu \vp_a \, H_a^0 + \eta \, Q_{\rm ppal} +
\nut \, C}. \lab{bf}
\ee
The model possesses six Dirac spinor fields $\psi^i$, and $\psit^i$,
$i=1,2,3$, which live in the subspaces with non-zero grades.
i.e. we have
\be
F^{\pm} \equiv E_{\pm 3} + F^{\pm}_{1} + F^{\pm}_{2}, \lab{gcon}
\ee
where $E_{\pm 3}$ are fixed elements of the subspaces $\gg_{\pm 3}$, and
the mappings $F^\pm_1$ and $F^\pm_2$ take values in the subspaces $\gg_{\pm
1}$ and $\gg_{\pm 2}$ respectively. We choose the elements $E_{\pm 3}$ as
\be
E_{\pm 3} = \frac{1}{6} \left[ \( 2 m_1 + m_2\) \,
H_1^{\pm 1} + \( 2 m_2 + m_1\)\, H_2^{\pm 1} \right],
\lab{epm3}
\ee
and use the following parametrisation for $F^\pm_1$ and $F^\pm_2$:
\br
F^+_{2} &=& \h\, \left[ m_3 \, \psi_R^3 \, E_{\a_3}^0 +
m_1 \, \psit_R^1 \, E_{-\a_1}^1 +
m_2 \, \psit_R^2 \, E_{-\a_2}^1\right],
\lab{fdef1}\\
F^+_{1} &=& \h\, \left[ m_1 \, \psi_R^1 \, E_{\a_1}^0 +
m_2 \, \psi_R^2 \, E_{\a_2}^0 +
m_3 \, \psit_R^3 \, E_{-\a_3}^1 \right], \\
F^-_{1} &=& \h\, \left[ m_3 \, \psi_L^3 \, E_{\a_3}^{-1} -
m_1 \, \psit_L^1 \, E_{-\a_1}^{0} -
m_2 \, \psit_L^2 \, E_{-\a_2}^{0} \right], \\
F^-_{2} &=& \h\, \left[ m_1 \, \psi_L^1 \, E_{\a_1}^{-1} +
m_2 \, \psi_L^2 \, E_{\a_2}^{-1} -
m_3 \, \psit_L^3 \, E_{-\a_3}^0 \right].
\lab{fdef2}
\er

The  model described by equations \rf{eqmot1}--\rf{eqmot2} is
invariant under the conformal transformations
\be
x_{+} \ra f(x_{+}) \, , \qquad x_{-} \ra g(x_{-}),
\lab{ct}
\ee
with the fields transforming as
\br
\vp_a (x_{+}\, , \, x_{-}) &\ra&
{\hat \vp}_a({\hat x}_{+}\, , \,  {\hat x}_{-}) = \vp_a (x_{+}\, , \,
x_{-})
\, ,
\\
\rme^{-{\widetilde   \nu} (x_{+}\, , \, x_{-})} &\ra& \rme^{-{\hat
{\widetilde \nu}}({\hat x}_{+}\, , \,
{\hat x}_{-})} = \( f^{\pr}\)^{\d} \, \( g^{\pr}\)^{\d}
\rme^{-{\widetilde   \nu} (x_{+}\, , \, x_{-})} \, ,
\\
\rme^{-\eta (x_{+}\, , \, x_{-})} &\ra& e^{-{\hat \eta}({\hat x}_{+}\, , \,
{\hat x}_{-})} = \( f^{\pr}\)^{1/3} \, \( g^{\pr}\)^{1/3}  e^{-\eta
(x_{+}\,
, \, x_{-})} \, ,
\lab{ctf}\\
\psi^i (x_{+}\, , \, x_{-}) &\ra & {\hat {\psi}}^i ({\hat x}_{+}\, , \,
{\hat x}_{-}) =   \rme^{\h \( 1+\gamma_5\)
\log\(f^{\pr}\)^{\frac{h\( \a_i\)}{3}-1}} \,
\rme^{\h \( 1-\gamma_5\)
\log\(g^{\pr}\)^{-\frac{h\( \a_i\)}{3}}} \, \psi^i (x_{+}\, , \, x_{-}),
\\
{\widetilde \psi}^i (x_{+}\, , \, x_{-}) &\ra & {\hat {{\widetilde
\psi}}}^i
({\hat x}_{+}\, , \,  {\hat x}_{-}) = \rme^{\h \( 1+\gamma_5\)
\log\(f^{\pr}\)^{-\frac{h\( \a_i\)}{3}}} \,
\rme^{\h \( 1-\gamma_5\)
\log\(g^{\pr}\)^{\frac{h\( \a_i\)}{3}-1}}
\, {\widetilde \psi}^i (x_{+}\, , \, x_{-}), \hspace{2em}
\er
where $h\( \a_i\)$ is the height of the root, i.e. $h\( \a_1\)=h\(
\a_i\)=1$ and $h\( \a_3\)=2$, and where the conformal weight $\d$,
associated to $\rme^{-\nu}$, is arbitrary. The two dimensional Lorentz
transformations $x_+ \to \l x_+$ and $x_- \to x_-/\l$ are contained in
\rf{ct}, as one observes by choosing $f(x_{+}) = \l \,
x_{+}$ and $g(x_{-}) = x_{-}/\l$.

Since $E_{\pm 3}$ commutes with the Cartan subalgebra generators $H_1$ and
$H_2$, it follows that equations \rf{eqmot1}--\rf{eqmot2} are also
invariant under the transformations
\br
B(x^+, x^-) &\ra& h_L(x^-) \, B(x^+, x^-) \, h_R(x^+), \lab{gs1} \\
F^+_m (x^+, x^-) &\ra& h_R^{-1}(x^+) \, F^+_m(x^+,x^-) \, h_R(x^+), \qquad
\lab{gs2}\\
F^-_m(x^+, x^-) &\ra& h_L(x^-) \, F^-_m(x^+, x^-) \, h_L^{-1}(x^-), \qquad
\lab{gs3}
\er
where
\be
h_L (x^-) = \rme^{\iu [\xi_L^1 (x^-) \, H_1 + \xi_L^2 (x^-) \, H_2]},
\qquad
h_R (x^+) = \rme^{\iu [\xi_R^1 (x^+) \, H_1 + \xi_R^2 (x^+) \, H_2]}.
\ee
That implies that the fields transform as
\br
\vp_a &\ra& \vp_a + \xi_L^a + \xi_R^a, \qquad \eta \ra \eta, \qquad
{\widetilde \nu} \ra {\widetilde   \nu}, \\
\psi^1 &\ra& \rme^{\frac{\mathrm i}{2} (1 + \gamma_5)(-2\xi_R^1 +
\xi_R^2)}\, \rme^{\frac{\mathrm i}{2}(1 - \gamma_5)(2\xi_L^1 -
\xi_L^2)}\,\psi^1, \\
\psi^2 &\ra& \rme^{\frac{\mathrm i}{2} (1 + \gamma_5)(\xi_R^1 -
2\xi_R^2)}\, \rme^{\frac{\mathrm i}{2} (1 - \gamma_5)(-\xi_L^1 +
2\xi_L^2)}\,\psi^2, \\
\psi^3 &\ra& \rme^{\frac{\mathrm i}{2} (1 + \gamma_5)(-\xi_R^1 -
\xi_R^2)}\,
\rme^{\frac{\mathrm i}{2}(1 - \gamma_5)(\xi_L^1+\xi_L^2)}\,\psi^3
\er
and ${\widetilde \psi}^i$ transforms in the same way as $\psi^i$ with
$\xi_{R,L}^a$ replaced by $-\xi_{R,L}^a$.

Notice that, by taking $\xi_R^a = - \xi_L^a = -\h \theta^a$, with
constant $\theta^a$, we get a $\mathrm{GL}_1(\bbC) \times
\mathrm{GL}_1(\bbC)$ global symmetry, with the fields transforming as
\be
\vp_a \ra \vp_a, \qquad \eta \ra \eta, \qquad {\widetilde   \nu} \ra
{\widetilde \nu}
\ee
and
\be
\psi^1 \ra \rme^{\iu (2\theta^1-\theta^2)}\psi^1, \qquad
\psi^2 \ra \rme^{\iu (2\theta^2-\theta^1)}\psi^2, \qquad
\psi^3 \ra \rme^{\iu (\theta^1+\theta^2)}\psi^2
\ee
with ${\widetilde \psi}^i$ transforming in the same way as $\psi^i$ with
$\theta^a$ replaced by $-\theta^a$.

On the other hand, if we take $\xi_R^a = \xi_L^a = -\h \chi^a$, with
constant $\chi^a$, we get a $\mathrm{GL}_1(\bbC) \times
\mathrm{GL}_1(\bbC)$ chiral symmetry,
\be
\vp_a \ra \vp_a -\chi^a, \qquad  \eta \ra \eta, \qquad {\widetilde \nu} \ra
{\widetilde \nu}
\ee
and
\be
\psi^1 \ra \rme^{\iu \gamma_5 ( 2\chi^1-\chi^2)}\psi^1, \qquad
\psi^2 \ra \rme^{\iu \gamma_5 ( 2\chi^2-\chi^1)}\psi^2, \qquad
\psi^3 \ra \rme^{\iu \gamma_5 ( \chi^1+\chi^2)}\psi^2
\ee
with ${\widetilde \psi}^i$ transforming in the same way as $\psi^i$ with
$\chi^a$ replaced by $-\chi^a$.

If our theory had a Lagrangian it would follow from the Noether theorem and
the above $\mathrm{GL}_1(\bbC) \times \mathrm{GL}_1(\bbC)$ symmetries that
it would possess two vector and two chiral  conserved currents. However, a
careful analysis of the equations of motion \rf{eqmot1}--\rf{eqmot2}
reveals that only half of such currents exist. Indeed, one can check that,
as a consequence of \rf{eqmot1}--\rf{eqmot2}, the vector current
\be
J_\mu = \iu \sum_{i=1}^3 m_i^2 \, {\bar \psi}^{i} \gamma_\mu
\psi^i
\lab{veccur}
\ee
and the chiral current
\be
J_{\mu}^5 = \iu \sum_{i=1}^3 m_i^2 \, {\bar \psi}^{i} \gamma_\mu \gamma_5
\psi^i + 2 \, \pa_{\mu} ( m_1 \vp_1 + m_2 \vp_2)
\lab{chicur}
\ee
are conserved
\be
\pa^{\mu} J_{\mu} = 0, \qquad \pa^{\mu} J_{\mu}^5 = 0.
\ee

The existence of these two conserved currents implies that the quantities
\be
{\cal J} = \iu \sum_{i=1}^3 m_i^2 \,\psit^i_R\psi^i_R +
2 \, \pa_{+} ( m_1 \vp_1 + m_2 \vp_2)
\ee
and
\be
{\bar {\cal J}} = \iu \sum_{i=1}^3 m_i^2 \,\psit^i_L\psi^i_L -
2 \, \pa_{-} \( m_1 \vp_1 + m_2 \vp_2\)
\ee
satisfy
\be
\pa_{-} {\cal J} = 0, \qquad \pa_{+}{\bar {\cal J}} = 0.
\lab{chiralcons}
\ee
Indeed, the currents $J_{\mu}^{(\pm)} = J_{\mu} \pm J^5_{\mu}$ are
obviously conserved. Their light front components have the form
\begin{eqnarray}
& J^{(+)}_{+} = 2 \, \iu \sum_{i=1}^3 m_i^2 \,\psit^i_R\psi^i_R +
2 \, \pa_{+} \( m_1 \vp_1 + m_2 \vp_2\), \quad J^{(+)}_{-} = 2 \, \pa_{-}\(
m_1 \vp_1 + m_2 \vp_2\), \\
& J^{(-)}_{+} = - 2 \, \pa_{+}\( m_1 \vp_1 + m_2 \vp_2\), \quad
J^{(-)}_{-} = 2 \, \iu \sum_{i=1}^3 m_i^2 \,\psit^i_L\psi^i_L -
2 \, \pa_{-} \( m_1 \vp_1 + m_2 \vp_2\). & \hspace{3em}
\end{eqnarray}
and the conservation laws
\begin{equation}
\pa_{+}J^{(\pm)}_{-}+\pa_{-}J^{(\pm)}_{+}=0
\end{equation}
lead to relations \rf{chiralcons}.

One can now perform a reduction of the theory by imposing the constraints
\be
{\cal J} = 0, \qquad {\bar {\cal J}} = 0,
\lab{constraints}
\ee
which are equivalent to\footnote{We use the normalisation
$\epsilon^{01}=1$.}
\be
\iu \sum_{i=1}^3 m_i^2 \,{\bar \psi}^i\gamma^{\mu}\psi^i =
-2 \, \epsilon^{\mu\nu}\pa_{\nu} \( m_1 \vp_1 + m_2 \vp_2\)
\lab{equivcur}
\ee
Consequently, in the submodel defined by \rf{constraints}, the vector
current \rf{veccur} is proportional to a topological current. The relation
\rf{equivcur} has very important consequences for the physical properties
of the theory. The time component of it implies that the charge density
associated to the current \rf{veccur} is proportional to the space
derivative of the $\vp$ fields, i.e.
\be
J^0 = \iu \sum_{i=1}^3 m_i^2 \,{\bar \psi}^i\gamma^{0}\psi^i =
-2 \, \pa_{x} ( m_1 \vp_1 + m_2 \vp_2 ).
\ee
Consequently, there can exist charges only on those regions where the space
derivative of the scalar fields are non-vanishing. As we will see below the
soliton solutions of this theory are localized in space, in the sense that
the scalar fields are not constant only in a region with a  size
determined by the soliton masses. Therefore, in the quantum theory, if we
look for excitations around a soliton solution, the spinor field can live
outside the solitons in configurations with charge zero only. We then have
a confining mechanism  which resembles that of the bag model for QCD.

\section{Dressing transformations and the Hirota method}
\label{sec:dressing}

We construct the soliton solutions using a combination of the so called
dressing transfomantion method \ct{dressing} and the Hirota method
\ct{hirota}. Let us start with the discussions of the former one.

The fact that the potentials $A_{\pm}$ in \rf{gp} satisfy the zero
curvature \rf{zc} implies that they can be represented as
\be
A_{\mu} = - \pa_{\mu} T \, T^{-1}, \lab{pvg}
\ee
where $T$ is a group element obtained by exponentiating the affine
Kac-Moody algebra $\asl3$ (see appendix \ref{sec:app}). In addition, the
potentials $A_\pm$ satisfy the so called grading condition of the
Leznov--Saveliev method \ct{ls}. This condition states that the potentials
$A_\pm$ has some limitations on their decomposition over grading
subspaces. In the case under consideration the potential $A_-$ has only
components belonging to the subspaces $\gg_0$, $\gg_{-1}$, $\gg_{-2}$ and
$\gg_{-3}$, and the potential $A_+$ has only components from the subspaces
$\gg_{+1}$, $\gg_{+2}$ and $\gg_{+3}$. It can be shown \ct{ls} that
potentials satisfying the grading condition can be represented in form
\rf{gp}. Here $E_{\pm 3}$ are abitrary mappings taking values in $\gg_{\pm
3}$ and satisfying the conditions
\be
\partial_+ E_{-3} = 0, \qquad \partial_- E_{+3} = 0.
\ee
Choosing them in form \rf{epm3} one fixes the class of
equations under
consideration. The dressing transformation method allows to construct gauge
transformations which do not violate the grading condition and the form of
$E_{\pm 3}$. Therefore one gets new solutions to the equations of motion
starting from some initial solution to these equations.

The concrete procedure to obtain desired gauge transformation is as
follows. Consider a constant group elements $\rho$ such that the element
\be
\Sigma = T \rho \, T^{-1}
\ee
admits the generalized Gauss decomposition
\be
\Sigma = \Sigma_- \, \Sigma_0 \, \Sigma_+, \lab{gauss}
\ee
where $\Sigma_-$, $\Sigma_0$, and $\Sigma_+$ take values in the subgroups
corresponding respectively to the subalgebras $\gg_{<0}$, $\gg_0$ and
$\gg_{>0}$ defined in appendix \ref{sec:app}. If decomposition \rf{gauss}
exists, we can introduce a new group element $T^\rho$ related to $T$ in two
different ways:
\be
T^\rho =  \Sigma_+ \, T = \Sigma_0^{-1} \, \Sigma_-^{-1} \, T \, \rho.
\ee
We now define the flat connection
\be
A_\mu^\rho = - \pa_\mu T^\rho \, {T^\rho}^{-1} = \Theta_\pm A_\mu
\Theta_\pm^{-1} - \pa_\mu \Theta_\pm \Theta_\pm^{-1}
\lab{potrho}
\ee
with
\be
\Theta_+ = \Sigma_+, \qquad  \Theta_- = \Sigma_0^{-1} \, \Sigma_-^{-1}
\equiv \Sigma_0^{-1} \, \widetilde \Theta_-.
\ee
We see that $A_{\mu}^\rho$ and $A_{\mu}$ are related by two gauge
tranformations; one with a group element involving only non-negative grade
generators and the other only non-positive grade generators. That fact
implies that $A_\mu^\rho$ satisfies the grading condition. Moreover, it can
be shown that the gauge transformations under consideration preserve the
form of $E_{\pm 3}$. Consequently, the dressing method defines
transformations on the space of solutions of the model defined by the zero
curvature condition \rf{zc}. Such transformations are parametrised by the
constant group element $\rho$.

A quite general procedure to construct soliton solutions in integrable
theories, using the dressing transformation method, is described in
paper \ct{tau}. It constitutes a generalization to practically any
two-dimensional integrable hierarchy, of the so-called ``solitonic
specialization''  in the context of the Leznov--Saveliev solution for Toda
type models \ct{otu,ous,ls}. The idea is to start from a ``vacuum''
solution  such that the connections $A_{\pm}$, when evaluated on it, belong
to an algebra of oscillators, i. e. an abelian (up to central terms)
subalgebra of the Kac--Moody algebra. One then looks for the eigenvectors
$V_i$, in $\cgh$, of such oscillators. The solitons belong to the orbits of
solutions obtained by the dressing transformation performed by elements of
the form $\rho = \rme^{V_{i_1}} \, \rme^{V_{i_2}} \ldots \rme^{V_{i_n}}$.

Notice that
\be
\widetilde \nu = -\frac{1}{12} \sum_{i=1}^3 m_i^2 \, x^+ x^-, \qquad \vp_a
= \eta = \psi^i = \psit^i = 0,\qquad a = 1, 2, \quad i = 1, 2, 3,
\ee
is a ``vacuum'' solution of the equations of motion
\rf{eqmot1}--\rf{eqmot2}. The potentials \rf{gp} evaluated on it become
\be
A_+ = - E_{+3}, \qquad A_- =  E_{-3} + \frac{1}{12} \sum_{i=1}^3 m_i^2 \,
x^+ \, C \lab{vacpot}
\ee
and so we can take the initial group element $T$ in the
form\footnote{Recall that the group element $T$ entering \rf{pvg} is
defined up to the right multiplication by a constant element of the group.}
\be
T = \rme^{x^+ E_{+3}} \, \rme^{-x^- E_{-3}}.
\lab{tvac}
\ee
Here we have used the equality
\be
\sbr{E_{+3}}{E_{-3}}= \frac{1}{12} \sum_{i=1}^3 m_i^2 \, C
\ee
which follows from \rf{epm3} and (\ref{a.3}).

Notice that $E_{\pm 3}$ belong to the ``algebra of oscillators'' generated
by the elements $H_a^n$ (see relation (\ref{a.3})). That is the so-called
homogeneous Heisenberg subalgebra \ct{kac} of the affine Kac--Moody algebra
$\asl3$.

The eigenvectors of $E_{\pm 3}$ are
\be
V_{\pm \a_i}(z) = \sum_{n=-\infty}^{\infty} z^{-n} E_{\pm \a_i}^n
\qquad i = 1, 2, 3,
\lab{vdef}
\ee
with $z$ being an arbitrary parameter. Indeed, one has
\be
\sbr{E_{+3}}{V_{\pm \a_i}(z)} = \pm \frac{z}{2} \, m_i V_{\pm \a_i}(z),
\qquad \sbr{E_{-3}}{V_{\pm \a_i}(z)} = \pm \frac{1}{2z} \, m_i V_{\pm
\a_i}(z).
\lab{eigenvectors}
\ee
Notice that $V_{\a_i}(z)$ and $V_{-\a_i}(-z)$ have the same eigenvalues. It
turns out that the soliton solutions are obtained by taking the constant
group element $\rho$ to be products of exponentials of the operators
\be
V_{\a_i}( a_{\a_i}^{\pm}, \, z) = a_{\a_i}^{+} V_{\a_i}(z) + a^{-}_{\a_i}
V_{-\a_i}(-z).
\lab{vadef}
\ee

We now perform the dressing transformation \rf{potrho} starting from the
vacuum potential \rf{vacpot}. The transformed potential $A_{\mu}^{\rho}$
has the structure given by relation \rf{gp} with $B$ and $F^\pm$ given by
\rf{bf} and \rf{gcon}, therefore we have the equality
\be
b ( E_{+3} + F_1^{+} + F_2^{+}) b^{-1} = \Theta_{\pm} E_{+3}
\Theta_{\pm}^{-1} + \pa_{+} \Theta_{\pm} \, \Theta_{\pm}^{-1},
\lab{dressplus}
\ee
where
\be
b = \rme^{\iu (\vp_1 \, H_1 + \vp_2 \, H_2)},
\ee
and the equality
\begin{eqnarray}
- \iu \pa_{-} (\vp_1 \, H_1 + \vp_2 \, H_2) - \pa_{-} {\widetilde \nu} \,
C &+& E_{-3} + F_1^{-} + F_2^{-} \nonu \\
&=& \Theta_{\pm} E_{-3}\Theta_{\pm}^{-1} - \pa_{-} \Theta_{\pm} \,
\Theta_{\pm}^{-1} + \frac{1}{12} \sum_{i=1}^3 m_i^2 \, x^+\, C.
\lab{dressminus}
\end{eqnarray}
These equalities relate the new fields and the parameters of the dressing
transformation. Notice that the $\eta$ field appears in \rf{gp} multiplying
the grading operator $Q_{\mathrm{ppal}}$, and that involves the operator
$D$ (see \rf{gradop}). Since $D$ is not the result of any commutator
it follows that the dressing transformation method does not excite the
field $\eta$ if one starts from a solution where it vanishes.

In order to construct the solution solutions we have to split
\rf{dressplus} and \rf{dressminus} into the grading subspaces. For
instance, taking grade $3$ component of \rf{dressplus} for $\Theta_{-}$,
and the grade $0$ one of \rf{dressminus} for $\Theta_{-}$ too, gives
\be
\Sigma_0^{-1} = b \, \rme^{\({\widetilde \nu} + \frac{1}{12}
\sum_{i=1}^3 m_i^2 \, x^+ x^-\) \, C}.
\ee
Then from \rf{gauss} and \rf{tvac} one gets for
\br
\Sigma = \rme^{x^+ E_{+3}} \, \rme^{-x^- E_{-3}}\, \rho \, \rme^{x^-
E_{-3}}\, e^{-x^+ E_{+3}} = {\widetilde \Theta}_{-}^{-1}\, b^{-1}  \,
\Theta_{+} \, e^{-\({\widetilde \nu} + \frac{1}{12} \sum_{i=1}^3 m_i^2 \,
x^+ x^-\) \, C}.
\lab{relforg}
\er

The idea now is to consider  matrix elements of both sides of \rf{relforg}
in states of the fundamental representations of $\cgh$ to obtain the
explict space-time dependence of the fields evaluated on the solutions
determined by the constant group element $\rho$. For instance, one obtains
from \rf{relforg} that
\be
\rme^{-\iu \vp_1} = \frac{\tau_1}{\tau_0}, \qquad \rme^{-\iu \vp_2} =
\frac{\tau_2}{\tau_0}, \qquad \rme^{-\({\widetilde \nu} + \frac{1}{12}
\sum_{i=1}^3 m_i^2 \, x^+ x^-\)} = \tau_0.
\lab{phitau}
\ee
with
\be
\tau_j = \bra{\l_j} \, \Sigma  \, \ket{\l_j}, \qquad \qquad j=0,1,2,
\lab{tauj}
\ee
where $\ket{\l_j}$ are the highest weight states of the fundamental
representations of $\cgh$, satisfying \rf{fundrep1} and \rf{fundrep2}, and
where we have used that, as a consequence of \rf{fundrep2} and
\rf{fundrep3},
\be
\Theta_+ \, \ket{\l_j} = \ket{\l_j}, \qquad \bra{\l_j} \, \widetilde
\Theta_-^{-1} = \bra{\l_j}. \lab{thetakill}
\ee

In order to obtain the expressions for the spinor fields we represent the
mappings $\Theta_+$ and $\widetilde \Theta^-$ as
\be
\Theta_+ = \rme^{\sum_{m>0} t^{(+m)}}, \qquad \widetilde \Theta_- =
\rme^{\sum_{m>0} t^{(-m)}},
\ee
where the mappings $t^{(+m)}$ and $t^{(-m)}$ take values in the subsapces
$\gg_{+m}$ and $\gg_{-m}$, respectively. According to \rf{fundrep2} and
\rf{fundrep3} for the highest weight vectors of the fundamental
reparesentations we have
\be
t^{(+m)} \,\ket{\l_j} = 0, \qquad \bra{\l_j} \, t^{(-m)} = 0, \qquad
m > 0, \qquad j = 0,1,2.
\ee
Considering the components   of grades $1$ and $2$ of \rf{dressplus} for
$\Theta_{-}$, implies that
\be
F_2^+ = [\,t^{(-1)}, \, E_{+3}\,], \qquad
F_1^{+} = \sbr{t^{(-2)}}{E_{+3}} + \h \, \sbr{t^{(-1)}} {\sbr{t^{(-1)}}
{E_{+3}}}.
\lab{frelt1}
\ee
Analogously, considering the components  of grades $-1$ and $-2$ of
\rf{dressminus} for $\Theta_{+}$, one gets
\be
F_2^{-} = \sbr{t^{(1)}}{E_{-3}}\; ; \qquad \qquad
F_1^{-} = \sbr{t^{(2)}}{E_{-3}} + \h \sbr{t^{(1)}}{\sbr{t^{(1)}}{E_{-3}}}.
\lab{frelt2}
\ee

Now, multiplying both sides of \rf{relforg} from the left by generators of
grades $1$ and $2$ and taking the expectation value on the states
$\ket{\l_j}$, one can relate the components of $t^{(-1)}$ and $t^{(-2)}$
to the matrix elements of  $\Sigma$. The peculiarity here is
that to extract  $t^{(-2)}$ one needs to consider two matrix elements  for
each one of its  components. By adding and subtracting the corresponding
relations coming from   \rf{relforg}, one obtains the components
of  $t^{(-2)}$ in terms of matrix elements of $\Sigma $, and a quadratic
relation involving the components of $t^{(-1)}$. A similar thing happens
when we multiply both sides of \rf{relforg} from the right by generators of
grades $-1$ and $-2$ to relate $t^{(+1)}$ and $t^{(+2)}$ to matrix elements
of $\Sigma $. Once we have obtained the relations between $t^{(\pm 1)}$ and
$t^{(\pm 2)}$ to matrix elements of $\Sigma $, we can use \rf{frelt1},
\rf{frelt2} and \rf{fdef1}--\rf{fdef2} to obtain the explict space-time
dependence of spinor fields on the solutions determined by $\rho$. The
results are
\br
{\psi}_R^1&=&
\frac{m_3}{m_1}\frac{\tseventeen}{\tthree}
-\frac{m_2}{m_1} \frac{\tsixteen}{\tone},
\qquad \qquad
{\psi}_L^1=-\frac{\tseven}{\ttwo},
\lab{taucampos1} \\
{\psi}_R^2&=&
\frac{m_3}{m_2}\frac{\tnineteen}{\ttwo}
-\frac{m_1}{m_2}\frac{\teighteen}{\tone},
\qquad \qquad
{\psi}_L^2=-\frac{\teight}{\tthree},
\\
{\psi}_R^3&=&\frac{\tsix}{\tone},
\qquad \qquad \qquad \qquad \qquad \quad \;
{\psi}_L^3=\frac{m_2}{m_3} \frac{\tfourteen}{\ttwo}
+\frac{m_1}{m_3}\frac{\tfifteen}{\tthree},
\\
{\tilde \psi}_R^1&=&-\frac{\tfour}{\ttwo},
\qquad \qquad \qquad \qquad \quad \quad \;\,
{\tilde \psi}_L^1=
\frac{m_2}{m_1}\frac{\televen}{\tone}
-\frac{m_3}{m_1}\frac{\tten}{\tthree},
\\
{\tilde \psi}_R^2&=&-\frac{\tfive}{\tthree},
\qquad \qquad \qquad \qquad \quad \quad \;\,
{\tilde \psi}_L^2=
\frac{m_1}{m_2}\frac{\tthirteen}{\tone}
-\frac{m_3}{m_2}\frac{\ttwelve}{\ttwo},
\\
{\tilde\psi}_R^3&=&
-\frac{m_2}{m_3}\frac{\ttwenty}{\ttwo}
-\frac{m_1}{m_3}\frac{\ttwentyone}{\tthree},
\qquad \quad \,
{\tilde \psi}_L^3=-\frac{\tnine}{\tone},
\lab{taucampos2}
\er
where $\tau_j$, $j=0,1,2$, are defined in \rf{tauj}, and where we have
denoted
\br
\tseventeen &=& \braopket{\l_2}{E_{-\a_1}^1 \Sigma}{\l_2}, \hspace{3em}
\tsixteen = \braopket{\l_0}{E_{-\a_1}^1 \Sigma}{\l_0}, \lab{taudef1} \\
\tseven &=& \braopket{\l_1}{\Sigma  E_{-\a_1}^0}{\l_1}, \hspace{3em}
\teighteen = \braopket{\l_0}{E_{-\a_2}^1 \Sigma}{\l_0}, \\
\tnineteen &=& \braopket{\l_1}{E_{-\a_2}^1 \Sigma}{\l_1}, \hspace{3.8em}
\teight = \braopket{\l_2}{\Sigma  E_{-\a_2}^0}{\l_2}, \\
\tsix &=& \braopket{\l_0}{E_{-\a_3}^1 \Sigma}{\l_0}, \hspace{3.1em}
\tfifteen = - \braopket{\l_2}{\Sigma  E_{-\a_3}^0}{\l_2}, \\
\tfourteen &=& - \braopket{\l_1}{\Sigma  E_{-\a_3}^0}{\l_1}, \hspace{3em}
\tfour = \braopket{\l_1}{E_{\a_1}^0 \Sigma}{\l_1}, \\
\tten &=& \braopket{\l_2}{\Sigma  E_{\a_1}^{-1}}{\l_2}, \hspace{3.5em}
\televen = \braopket{\l_0}{\Sigma  E_{\a_1}^{-1}}{\l_0}, \\
\tfive &=& \braopket{\l_2}{E_{\a_2}^0 \Sigma}{\mid \l_2}, \hspace{2.8em}
\ttwelve = \braopket{\l_1}{\Sigma  E_{\a_2}^{-1}}{\l_1}, \\
\tthirteen &=& \braopket{\l_0}{\Sigma  E_{\a_2}^{-1}}{\l_0}, \hspace{3.5em}
\ttwentyone = \braopket{\l_2}{E_{\a_3}^0 \Sigma}{\l_2}, \\
\ttwenty &=& \braopket{\l_1}{E_{\a_3}^0 \Sigma}{\l_1}, \hspace{4.4em}
\tnine = \braopket{\l_0}{\Sigma  E_{\a_3}^{-1}}{\l_0}. \lab{taudef2}
\er

As we mentioned above in the process of obtaining the components of
$t^{(\pm 2)}$ in terms of the matrix elements of $\Sigma$ one gets
quadratic relations involving the components of $t^{(\pm 1)}$. These
relations are given by
\br
\tfive\,\tsix + \tthree\,\tsixteen - \tone\,\tseventeen &=& 0, \qquad
\tfour\,\tsix - \ttwo\,\teighteen +\tone\,\tnineteen = 0,
\lab{algrel1} \\
\tfour\,\tfive - \tthree\,\ttwenty + \ttwo\,\ttwentyone &=& 0, \qquad
\teight\,\tnine - \tone\,\tten + \tthree\,\televen =0, \\
\tseven\,\tnine + \tone\,\ttwelve - \ttwo\,\tthirteen &=& 0, \qquad
\tseven\,\teight + \tthree\,\tfourteen - \ttwo\,\tfifteen = 0.
\lab{algrel2}
\er

According to the ``solitonic specialization'' \ct{tau,otu,ous,ls}, the
soliton solutions are obtained by taking $\rho$ to be the product of
exponentials of the operators \rf{vadef}. The one-soliton solutions are
obtained by taking
\be
\rho = \rho_i = \exp[V_{\a_i} (a_{\a_i}^{\pm},z)], \qquad i = 1,2,3.
\lab{ros}
\ee
Therefore, there are three species of solitons. Similarly, the two-soliton
solutions are obtained  by taking
\be
\rho = \rho_i \, \rho_j = \exp [V_{\a_i} (a_{\a_i}^{\pm}, z_1)] \, \exp
[V_{\a_j} (b_{\a_j}^{\pm}, z_2)],
\ee
and so there are six of such type of solutions. Using \rf{eigenvectors} one
then gets from \rf{relforg} that for the one-soliton solutions
\be
\Sigma = \Sigma _i = \rme^{x^+ E_{+3}} \, \rme^{-x^- E_{-3}}\, \rho_i \,
\rme^{x^- E_{-3}}\, \rme^{-x^+ E_{+3}} = \exp \left[ \rme^{\G _i (z)}
V_{\a_i} (a_{\a_i}^{\pm}, z) \right],
\lab{gforsol1}
\ee
and for the two-soliton solutions
\br
\Sigma = \Sigma _{ij} =  \Sigma _i \, \Sigma _j &=& \rme^{x^+ E_{+3}} \,
\rme^{-x^- E_{-3}} \, \rho_i \, \rho_j \, \rme^{x^-  E_{-3}}\, \rme^{-x^+
E_{+3}}
\nonu \\
&=& \exp \left[\rme^{\G_i(z_1)} V_{\a_i} ( a_{\a_i}^{\pm}, z_1) \right] \,
\exp \left[ \rme^{\G_j (z_2)} V_{\a_j} ( b_{\a_j}^{\pm}, z_2 )\right],
\lab{gforsol2}
\er
where
\be
\G_i\(z\) = \h \, m_i \( z x^+ - \frac{x^-}{z}\) = \g \( x - v t\)
\lab{biggammadef}
\ee
with
\be
\g = \h m_i \( z + \frac{1}{z}\)= ( {\rm sign}\; z)
\frac{m_i}{\sqrt{1-v^2}}
= \epsilon \,  m_i \, \cosh \theta
 \; ; \qquad  v = \frac{z^2-1}{z^2+1} = \tanh \theta,
\lab{gammadef}
\ee
where we have introduced the rapidity $\theta$ as $z= \epsilon \,
\rme^{\theta}$, with $\epsilon = \pm 1$.  Therefore, if $z$ is real we have
$|v| < 1$, where the speed of light has been normalized to unity. A
generalization to the $n$-soliton case is evident.

In order to obtain the final expressions for one-soliton and two-soliton
solutions one has to evaluate the matrix elements in \rf{tauj} and
\rf{taudef1}--\rf{taudef2} for the group elements \rf{gforsol1} and
\rf{gforsol2}. That calculation can be performed using vertex operator
representations for $\cgh$ \ct{goddoliv,kac}. We have to use in fact the
homogeneous vertex operator realization of the three fundamental
representations of $\cgh$. It turns out  that such representations are
integrable in the sense that the step operators of $\cgh$ are nilpotent
\ct{kac}. Indeed, one can verify that the operators \rf{vdef} satisfy
\ct{goddoliv,kac}
\be
V_{\a_i} \( z_1 \) V_{\a_i} \( z_2 \) \ra 0 \qquad {\rm as} \quad
z_1\ra z_2. \lab{nil}
\ee
That implies that the matrix elements in \rf{tauj} and
\rf{taudef1}--\rf{taudef2} truncate when one expands the exponentials in
\rf{gforsol1} and \rf{gforsol2}, and consequently they become polinomials
in the $\rme^{\G_i}$'s. In this sense, those matrix elements are the
Hirota tau-functions \ct{hirota} for the model described by equations
\rf{eqmot1}--\rf{eqmot2}. Note that due to \rf{nil} the maximal power of
each $\rme^{\Gamma_i}$ in the expansion of a tau-function over them is
equal to 2. Therefore, a general tau-function for an $n$-soliton solution
has the form
\be
\tau = \sum_{p_1, \ldots, p_n = 0}^2 c_{p_1 \ldots p_n} \exp \,
[p_1 \Gamma_{i_1}(z_1) + \cdots + p_n \Gamma_{i_n}(z_n)]. \lab{gt}
\ee

The connection between  dressing transformations, solitonic specialization
and the Hirota method has been discussed in \ct{tau} for any hierarchy of
integrable models possessing a zero curvature representation in terms of an
affine Kac--Moody algebra. It is a quite general result that explains a lot
of the structures known in soliton theory. In addition, on the practical
side it allows one to use the good features of each method.

The Hirota method is quite powerful  to actually evaluate the explicit
expression for the soliton solutions. Besides, being a recursive method it
allows a simple implementation on a computer program for algebraic
manipulation like {\sc Mathematica}. However, one of its difficulty is to
find the relation between the tau-functions and the fields that lead to the
truncation of the Hirota expansion. But the dressing transformation
method together with the solitonic specialization does exactly that.

The dressing method on the other hand, requires the evaluation of matrix
elements in vertex operator representations of the Kac-Moody algebra, which
in the case of higher soliton solutions become extremely tedious. One can
take advantage of those features to speed up calculations.

To use the Hirota method we have to find the equations satisfied by
tau-functions. They are determined by substituting the relations between
the fields and tau-functions \rf{phitau} and
\rf{taucampos1}--\rf{taucampos2} into the equations of motion
\rf{eqmot1}--\rf{eqmot2}. Notice that, except for $\eta$, we have
$15$ fields in the model  \rf{eqmot1}--\rf{eqmot2} (three scalars and six
two-component spinors). However, in \rf{phitau} and
\rf{taucampos1}--\rf{taucampos2} we have introduced $21$ tau-functions. The
six equations missing are the algebraic relations
\rf{algrel1}--\rf{algrel2} provided by the dressing method.

Substituting \rf{phitau} and \rf{taucampos1}--\rf{taucampos2} into
\rf{eqmot1}--\rf{eqmot2} one gets that the equations for $\vp_a$ and
${\widetilde \nu}$ lead to the following three Hirota equations
\br
4\,\ttwo\, \tthree\, (\tone\, \pa_{+}\pa_{-}\tone
-  \pa_{-}\tone\, \pa_{+}\tone) &=&
{{m_1}^2}\,\tone\,\ttwo\,\tfour\,\tseven
+ {{m_2}^2}\,\tone\,\tthree\,\tfive\,\teight \nonu \\
&-& {{m_2}^2}\,{{\tthree}^2}\,\tfourteen\,\ttwenty
- m_1\,m_2\,\ttwo\,\tthree\,\tfifteen\,\ttwenty \nonu \\
&-& m_1\,m_2\,\ttwo\,\tthree\,\tfourteen\,\ttwentyone
- {{m_1}^2}\,{{\ttwo}^2}\,\tfifteen\,\ttwentyone, \\
4\,\tone\,\tthree (\ttwo\,\pa_{+}\pa_{-}\ttwo -
 \pa_{-}\ttwo\,\pa_{+}\ttwo) &=&
{{m_2}^2}\,\ttwo\,\tthree\,\tfive\,\teight
+ {{m_3}^2}\,\tone\,\ttwo\,\tsix\,\tnine \nonu \\
&-& m_2\,m_3\,\tone\,\tthree\,\tten\,\tsixteen
+ {{m_2}^2}\,{{\tthree}^2}\,\televen\,\tsixteen\nonu\\
&+& {{m_3}^2}\,{{\tone}^2}\,\tten\,\tseventeen
- m_2\,m_3\,\tone\,\tthree\,\televen\,\tseventeen, \\
4\,\tone\,\ttwo (\tthree\,\pa_{+}\pa_{-}\tthree -
\pa_{-}\tthree\,\pa_{+}\tthree) &=&
{{m_1}^2}\,\ttwo\,\tthree\,\tfour\,\tseven
+ {{m_3}^2}\,\tone\,\tthree\,\tsix\,\tnine \nonu \\
&-& m_1\,m_3\,\tone\,\ttwo\,\ttwelve\,\teighteen
+ {{m_1}^2}\,{{\ttwo}^2}\,\tthirteen\,\teighteen \nonu\\
&+& {{m_3}^2}\,{{\tone}^2}\,\ttwelve\,\tnineteen
- m_1\,m_3\,\tone\,\ttwo\,\tthirteen\,\tnineteen.
\er
The components of the equations of motion \rf{eqmot1}--\rf{eqmot2} for
$\psi$'s and $\psit$'s that do not involve the quadratic terms $U^i$ and
${\widetilde U}^i$ (see \rf{zerous}) lead to the Hirota equations
\br
2\,\( \ttwo\,\pa_{+}\tseven - \tseven\,\pa_{+}\ttwo \) &=&
m_3\,\tone\,\tseventeen - m_2\,\tthree\,\tsixteen,
\\
2\,\( \tthree\,\pa_{+}\teight - \teight\,\pa_{+}\tthree \) &=&
m_3\,\tone\,\tnineteen - m_1\,\ttwo\,\teighteen,
\\
2\,\( \tone\,\pa_{-}\tsix - \tsix\,\pa_{-}\tone \) &=&
m_2\,\tthree\,\tfourteen + m_1\,\ttwo\,\tfifteen,
\\
2\,\( \ttwo\,\pa_{-}\tfour - \tfour\,\pa_{-}\ttwo \) &=&
m_3\,\tone\,\tten   - m_2\,\tthree\,\televen,
\\
2\,\( \tthree\,\pa_{-}\tfive - \tfive\,\pa_{-}\tthree \) &=&
m_3\,\tone\,\ttwelve   - m_1\,\ttwo\,\tthirteen,
\\
2\,\(\tone\,\pa_{+}\tnine - \tnine\,\pa_{+}\tone \) &=&
- m_2\,\tthree\,\ttwenty   - m_1\,\ttwo\,\ttwentyone,
\er
and the other components of the equations \rf{eqmot1}--\rf{eqmot2} for
$\psi$'s and $\psit$'s involving  the quad\-ra\-tic terms $U^i$ and
${\widetilde U}^i$ lead to
\br
&& 2 \, m_1\, \ttwo^2 \, (\tthree\,\pa_{+} \tfifteen - \tfifteen \, \pa_{+}
\tthree) + 2 \, m_2 \, \tthree^2 \,( \ttwo \, \pa_{+} \tfourteen -
\tfourteen \, \pa_{+} \ttwo) \nonu \\*
&& \hspace{2em} {} = m_2^2 \, \tthree^2 \, \teight \, \tsixteen
- m_3^2 \, \tone \, \ttwo \, \tthree \, \tsix - m_2 \, m_3 \, \tone \,
\tthree \, \teight \, \tseventeen \nonu \\*
&& \hspace{2em} {} - m_1^2 \, \ttwo^2 \, \tseven \, \teighteen + m_1 \, m_3
\, \tone \, \ttwo \, \tseven \, \tnineteen, \\[0.5em]
&& 2 \, m_2 \, \tthree^2 \, (\tone \, \pa_{-} \tsixteen - \tsixteen \,
\pa_{-} \tone) + 2 \, m_3 \, \tone^2 \, (\tseventeen \, \pa_{-} \tthree
- \tthree \, \pa_{-} \tseventeen) \nonu \\*
&& \hspace{2em} {} = m_1^2 \, \tone \, \ttwo \, \tthree \, \tseven
- m_3^2 \, \tone^2 \, \tsix \, \ttwelve + m_1 \, m_3 \, \tone \, \ttwo \,
\tsix \, \tthirteen \nonu \\*
&& \hspace{2em} {} - m_2^2 \, \tthree^2 \, \tfive \, \tfourteen - m_1 \,
m_2 \, \ttwo \, \tthree \, \tfive \, \tfifteen,
\\[0.5em]
&& 2 \, m_1 \, \ttwo^2 \, (\tone\,\pa_{-}\teighteen - \teighteen \, \pa_{-}
\tone) - 2 \, m_3 \, \tone^2 \, (\ttwo \, \pa_{-} \tnineteen - \tnineteen
\, \pa_{-} \ttwo) \nonu \\*
&& \hspace{2em} {} = m_2^2 \, \tone \, \ttwo \, \tthree \, \teight + m_3^2
\, \tone^2 \, \tsix \, \tten - m_2 \, m_3 \, \tone \, \tthree \, \tsix \,
\televen \nonu \\*
&& \hspace{2em} {} + m_1 \,m_2 \, \ttwo \, \tthree \, \tfour \, \tfourteen
+ m_1^2 \, \ttwo^2 \, \tfour \, \tfifteen, \\[0.5em]
&& 2 \, m_2 \, \tthree^2 \, (\tone \, \pa_{+} \televen - \televen \,
\pa_{+} \tone) + m_3 \, \tone^2 \, (\tten \, \pa_{+} \tthree - \tthree \,
\pa_{+} \tten) \nonu \\*
&& \hspace{2em} {} = m_1^2 \, \tone \, \ttwo \, \tthree \, \tfour + m_1 \,
m_3 \, \tone \, \ttwo \, \tnine \, \teighteen - m_3^2 \, \tone^2 \, \tnine
\, \tnineteen \nonu \\*
&& \hspace{2em} {} + m_2^2 \, \tthree^2 \, \teight \, \ttwenty + m_1 \, m_2
\, \ttwo \, \tthree \, \teight \, \ttwentyone, \\[0.5em]
&& 2 \, m_1 \, \ttwo^2 \, (\tone \, \pa_{+} \tthirteen - \tthirteen \,
\pa_{+} \tone) + m_3 \, \tone^2 \, (\ttwelve \, \pa_{+} \ttwo - \ttwo \,
\pa_{+} \ttwelve) \nonu \\*
&& \hspace{2em} {} = m_2^2 \, \tone \, \ttwo \, \tthree \, \tfive - m_2 \,
m_3 \, \tone \, \tthree \, \tnine \, \tsixteen + m_3^2 \, \tone^2 \, \tnine
\, \tseventeen \nonu \\*
&& \hspace{2em} {} - m_1 \, m_2 \, \ttwo \, \tthree \, \tseven \, \ttwenty
- m_1^2 \, \ttwo^2 \, \tseven \, \ttwentyone, \\[0.5em]
&& 2 \, m_1 \, \ttwo^2 \, (\tthree \, \pa_{-} \ttwentyone - \ttwentyone \,
\pa_{-} \tthree) + m_2 \, \tthree^2 \, (\ttwo \, \pa_{-} \ttwenty -
\ttwenty \, \pa_{-} \ttwo) \nonu \\*
&& \hspace{2em} {} = m_3^2 \, \tone \, \ttwo \, \tthree \, \tnine + m_2 \,
m_3 \, \tone \, \tthree \, \tfive \, \tten - m_2^2 \, \tthree^2 \, \tfive
\, \televen \nonu \\*
&& \hspace{2em} {} - m_1 \, m_3 \, \tone \, \ttwo \, \tfour \, \ttwelve +
m_1^2 \, \ttwo^2 \, \tfour \, \tthirteen.
\er

The equivalence between the vector and topological currents \rf{equivcur}
in terms of the tau-functions become
\br
&& 2\, m_1\,\tthree \, (\ttwo \,\pa_{+}\tone
- \tone \,\pa_{+}\ttwo) + 2 \, m_2\,\ttwo \, ( \tthree \,\pa_{+}\tone
- \tone \,\pa_{+}\tthree )
\nonu \\*
&& \hspace{2em} {}= m_1\,m_2\,\tthree \,\tsixteen \,\tfour
- m_1\,m_3\,\tone \,\tseventeen \,\tfour
+ m_1\,m_2\,\ttwo \,\teighteen \,\tfive
\nonu \\*
&& \hspace{2em} {} - m_2\,m_3\,\tone \,\tnineteen \,\tfive
- m_2\,m_3\,\tthree \,\tsix \,\ttwenty
- m_1\,m_3\,\ttwo \,\tsix \,\ttwentyone,
\lab{equivcurtau1}\\[0.5ex]
&& 2\, m_1\,\tthree \, ( \ttwo \,\pa_{-}\tone
- \tone \,\pa_{-}\ttwo ) + 2 \, m_2\,\ttwo \, ( \tthree \,\pa_{-}\tone
- \tone \,\pa_{-}\tthree )
\nonu\\*
&& \hspace{2em} {} = m_2\,m_3\,\tthree \,\tfourteen \,\tnine
+ m_1\,m_3\,\ttwo \,\tfifteen \,\tnine
+ m_1\,m_2\,\tthree \,\tseven \,\televen
\nonu\\*
&& \hspace{2em} {} - m_1\,m_3\,\tone \,\tseven \,\tten
+ m_1\,m_2\,\ttwo \,\teight \,\tthirteen
- m_2\,m_3\,\tone \,\teight \,\ttwelve.
\lab{equivcurtau2}
\er

Resuming we can say that the soliton solution can then be obtained by
evaluating the matrix elements \rf{tauj} and \rf{taudef1}--\rf{taudef2} in
the homogeneous vertex operator realization of the three fundamental
representations of the affine Kac-Moody algebra $\asl3$. Alternatively one
can obtain the same solutions by using  the Hirota method. In such case one
uses the ansatz \rf{gt} for the tau-function provided by the dressing
method. The coefficients $c_{p_1 \ldots p_n}$ can be determined recursively
using a program for algebraic manipulation. In appendix \ref{sec:app2} we
give the code of a {\sc Mathematica} program implementing the Hirota
method. The explicit soliton solutions are given in Sections
\ref{sec:onesol}
and \ref{sec:twosol}.

\sect{Topological charge}
\label{sec:topologicalcharge}

In $(1+1)$ dimensions the topological current is defined as $J^{\rm
top}_{\mu} \sim \epsilon_{\mu\nu} \pa^{\nu} \Phi$, where $\Phi$ is some
Lorentz scalar quantity. The reason is that if $\Phi$ does not present
singularities, the space-time derivatives acting on it commute and
therefore $J^{\rm top}_{\mu}$ is conserved independently of the
equations of motion, i.e. $\pa^{\mu}J^{\rm top}_{\mu} = 0$. The topological
charge is then $Q^{\rm top} = \int \mathrm d x \, J_{0}^{\rm top} \sim
\Phi(+\infty) - \Phi( -\infty)$. For finite energy solutions the
fields have to approach a vacuum configuration for $x\ra \pm \infty$, and
therefore the introduction of a topological current only makes sense in
theories with degenerate vacua. Our theory \rf{eqmot1}--\rf{eqmot2} has
four scalars
fields from which we can build topological charges. However, $\eta$ is a
free
field and its vacuum configuration do not lead to interesting charges. The
field ${\widetilde \nu}$ is not suitable either because one needs
${\widetilde
\nu} \ra  - \frac{1}{12}\, \sum_{i=1}^3 m_i^2 x^+ x^-$ as
$x\ra \pm\infty$. Therefore, we are
left with the two scalars $\vp_a$, $a=1,2$. Then, following the approach
adopted in the abelian affine Toda models we introduce the quantity
\be
\vp = \sum_{a=1}^2 \frac{2 \a_a}{\a_a^2} \, \vp_a,
\ee
where $\a_a$, $a=1,2$, are the simple roots of $\mathfrak{sl}_3(\bbC)$ (see
appendix \ref{sec:app}). We then have that $\vp_a = (\vp|\l_a)$, where
$\l_a$ are the fundamental weights of $\mathfrak{sl}_3(\bbC)$ defined by
the relation \ct{hump}
\be
2 \, \frac{(\a_a | \l_b)}{(\a_a | \a_a)} = \d_{ab}.
\ee
One observes from \rf{nonzerou1}--\rf{potentials} that the fields $\vp_a$
enter into the equations of motion through the combinations $\rme^{\pm i\,
(\vp | \a_j)}$, $j = 1, 2, 3$, which are invariant under the
transformations
\be
\vp \ra \vp + 2\pi \mu
\ee
where $\mu$ is any vector on the root space satisfying the condition $(\a_i
| \mu) \in \bbZ$. Such vectors are the so-called weights of
$\mathfrak{sl}_3(\bbC)$ and they constitute an infinite discrete lattice
called the weight lattice \ct{hump}. In the case of the abelian affine Toda
models the vacua of the theory is determined by such weight lattice.
However, in the model described by equations \rf{eqmot1}--\rf{eqmot2} one
notices that for any constants $\vp_a^{(0)}$ and $\eta^{(0)}$
\be
\psi^{i} = \psit^{i} = 0, \qquad \eta = \eta^{(0)}, \qquad \vp_a =
\vp_a^{(0)}, \qquad {\widetilde \nu} = - \frac{1}{12}\, \sum_{i=1}^3 m_i^2
\, x^+ x^-
\ee
is a vacuum configuration. Consequently, the vacuum configurations of the
$\vp_a$ fields are not determined by the weight lattice of
$\mathfrak{sl}_3(\bbC)$. What we will see below is that the topological
charges of the one-soliton solutions of \rf{eqmot1}--\rf{eqmot2} lie in
fact on the weight lattice of the three $\mathfrak{sl}_2(\bbC)$
subalgebras associated to the three positive roots of
$\mathfrak{sl}_3(\bbC)$.

We shall define the topological current and charge as
\be
J_\mu^{\mathrm{top}} = \frac{1}{2\pi} \, \epsilon_{\mu\nu} \pa^{\nu} \vp,
\qquad Q^{\mathrm{top}} = \int \mathrm d x \, J_{0}^{\rm top} =
\frac{1}{2\pi} \, [\vp (+\infty) - \vp ( -\infty)].
\lab{topcurchar}
\ee
In terms of the tau-functions introduced in \rf{phitau} one can express the
topological charge as
\be
Q^{\rm top} = \frac{\mathrm i}{2\pi}\, \sum_{a=1}^2 \frac{2 \a_a}{\a_a^2}
\, \ln \frac{\tau_a}{\tau_0} \, \bigg|_{-\infty}^{+\infty}.
\lab{tautopchar}
\ee
The relation \rf{phitau} implies that
$-\iu \vp_a = \ln |\tau_a/\tau_0| + \iu \mathrm{arg}(\tau_a/\tau_0)$.
Therefore, in order to have real solutions for the fields $\vp_a$, one
needs $|\tau_a| = |\tau_0|$. Given the solutions for the tau-functions, the
relation \rf{phitau} determines $\vp_a$ only modulo $2\pi$. Therefore, the
topological charges are defined up to an integer linear combination of the
simple roots $\a_a$ (recal that we use the normalization $\a_a^2=2$).
When evaluating the charges we should then use the prescription that the
arguments of the tau-functions lie between $\pi$ and $-\pi$.

Notice that the topological current in \rf{equivcur}, which is equivalent
to the spinor vector current is the projection of \rf{topcurchar} onto the
vector $-4\pi \( m_1 \l_1 + m_2 \l_2\)$.

\sect{Masses of fundamental particles and solitons}
\label{sec:masses}

The model described by equations \rf{eqmot1}--\rf{eqmot2} is conformally
invariant. It means, in particular, that if one has a solution of the
equations with some finite mass, then by a scale transformation it can be
continuosly transformed to a solution with any other mass. Hence to have
stable solutions we have to break the conformal invariance. The simplest
way
to do this is to freeze the field $\eta$. This fixes also the masses of the
fundamental fields. Indeed, taking, for example, $\eta = 0$, and
considering the linear part of the equations \rf{eqmot1}--\rf{eqmot2} one
observes that the masses of the fundamental particles are (see section $7$
of \ct{matter})
\be
m_{\psi^i} = m_{\psit^i} = m_i, \qquad m_{\vp_a} = m_{{\widetilde \nu}} =
0.
\ee

The mass of a soliton is defined to be its rest frame energy. The energy is
measured by the Hamiltonian or the energy-momentum tensor. The model under
consideration does not possess a Lagrangian and so we cannot calculate the
canonical energy-momentum tensor. However, in reference \ct{matter} it was
shown that our model belongs to a class of theories obtained by the
Hamiltonian reduction from the two-loop WZNW model \ct{twoloop,schwi}.
The energy-momentum tensor $L_{\mu\nu}$, of the two-loop WZNW model is of
the Sugawara form, i.e. quadratic in the conserved currents. We take the
energy momentum tensor of our model to be the reduced form of $L_{\mu\nu}$
under the Hamiltonian reduction mentioned above. Denote that by
$L_{\mu\nu}^{\mathrm{red}}$. Such tensor is conserved and traceless, i.e.
\ct{matter}
\be
\pa^{\mu} L_{\mu\nu}^{{\mathrm{red}}} = 0, \qquad \qquad
{L^{\mathrm{red}}_{\mu}}^{\mu} = 0.
\ee
However, according to the arguments given in \ct{hirnosso,matter} the
masses measured by such tensor must vanish, since it corresponds to a
conformally invariant field theory. On the other hand, the conformal
symmetry can be broken by freezing the $\eta$ field to a constant value. We
define the energy momentum tensor of such theory with a broken conformal
symmetry by
\be
\Theta_{\mu\nu}^{{\rm broken}} \equiv \Theta_{\mu\nu}\mid_{\eta =0},
\ee
where
\be
 \Theta_{\mu\nu} = L_{\mu\nu}^{\mathrm{red}} + S_{\mu\nu}
\lab{improvtensor}
\ee
with
\be
S_{\mu\nu} = \kappa ( \pa_{\mu}\pa_{\nu} - g_{\mu\nu} \pa^2 ) \left[
{\widetilde \nu} + \frac{\mathrm i}{3} \( \vp_1 +\vp_2\) \right]
\lab{smunu}
\ee
and $\kappa$ being the coupling constant of the two-loop WZNW model
\ct{matter,twoloop,schwi}. The tensor $\Theta_{\mu\nu}$ is conserved but
not traceless. We  define the soliton masses to be the space
integral of $\Theta_{00}^{{\rm broken}}$ on the soliton rest frame.

Therefore, using \rf{improvtensor} and \rf{smunu}, we obtain
the following expression for the mass $M$ of the soliton moving with the
velocity $v$:
\br
\frac{M}{\sqrt{1-v^2}} &=& - \int_{-\infty}^{\infty} \mathrm d x \,
(\Theta_{00}^{{\rm broken}} - E^{\mathrm{vac}}) \nonu \\
&=& - \kappa \, \pa_x \left[\widetilde \nu + \frac{\mathrm i}{3} \( \vp_1
+\vp_2\) + \frac{x^+ x^-}{12}\sum_{i=1}^3 m_i^2 \right]
\Bigg|_{-\infty}^{+\infty} = \frac{\kappa}{3} \, \pa_x \sum_{j=0}^2 \ln
\tau_j \, \bigg|_{-\infty}^{+\infty}.
\lab{solmass}
\er
Here we have used relation \rf{phitau}, and the fact that the integral of
$L_{\mu\nu}^{{\rm red.}}$ vanishes, as discussed above, due to the
conformal
symmetry. The term
$E^{\mathrm{vac}}$ is due to the nonzero vacuum energy density,
associated to the fact that the vacuum configuration of the field
$\widetilde \nu$ is $-(1/12)\sum_{i=1}^3 m_i^2 x^+ x^-$. Thus, the soliton
mass is determined by the asymptotic behaviour of the solution.

\sect{Time delays}
\label{sec:timedelays}

A soliton is a classical solution that travels with a constant speed
without dispersion and keeps its form under scattering on other solitons.
The effect of scattering is only a phase shift or a displacement in the
soliton position. Now we will show that the solitons we have been working
with are true solitons in this sense and calculate the so-called time
delays of the scattering of two solitons, using the procedure elaborated in
papers \ct{fring,su2matter}.

Consider two solitons that in the distant past are well apart, then
colide near $t = 0$ and then separate again in the distant future.
Therefore, except for the region where the scattering occurs, the solitons
are free and so travel with constant velocities. Let us write the
trajectories of one of the solitons before and after the collision as
\be
x = v t + x(I) \qquad {\rm and} \qquad x = v t + x(F),
\ee
since the velocity is the same. The lateral displacement at fixed time is
measured by
\be
\Delta(x) = x(F) - x(I), \lab{latdisp}
\ee
and the time delay is defined as
\be
\Delta(t) = t(F) - t(I) = -\frac{\Delta (x)}{v}
\lab{timedelay}
\ee
with the intercepts of the trajectories with the time axis being given by
$t(F)= - x(F)/v$ and $t(I)=-x(I)/v$. The lateral displacement and the time
delay are not Lorentz invariant, and so we consider the invariant
\be
E \Delta (x) = - p \Delta(t)
\lab{invdelta}
\ee
with $E$ and $p=v E$ being the energy and momentum of the soliton
respectively. Since $E$ is positive it follows that $\Delta (x)$ has the
same sign in any reference frame, with only its value being frame
dependent. The time delay on the other hand may change the sign under
Lorentz transformations. One can show that the lateral displacements and
time delays for the two solitons participating in the collision have to
satisfy \ct{fring,su2matter}
\be
E_1 \Delta_1 (x) + E_2 \Delta_2 (x) = 0, \qquad
p_1 \Delta_1(t) + p_2 \Delta_2(t) = 0
\lab{invdelay}
\ee
where the index $i$ labels the quantities associated to the $i$-th
particle. Notice therefore that, since the energies are positive, the
lateral displacements have opposite signs. Clearly, in the center of
momentum frame where $p_1 + p_2 = 0$, the time delays are equal, i.e.
$\Delta_1^{\rm cm} (t)=\Delta_2^{\rm cm}(t)$. Therefore, from
\rf{invdelta} we have that  $E_1^{\rm cm} \Delta_1^{\rm cm}(x)/p_1
= - E_2^{\rm cm} \Delta_2^{\rm cm}(x)/p_1 = -\Delta_1^{\rm cm}(t)
=-\Delta_2^{\rm cm}(t)$. Consequently, since $\Delta (x)$ has the same
sign in any frame,  it follows that, if the first particle moves to the
right faster then the second particle (so that $p_1$ is positive in the
center of momentum frame), then $(-\Delta_1 (x))$, $\Delta_2(x)$,
$\Delta_1^{\rm cm}(t)$ and $\Delta_2^{\rm cm}(t)$ all have the same sign.
The physical interpretation of that sign is related to the character
(attractive or repulsive) of the interaction forces. Indeed, if the
force is attractive then the first particle will accelerate as it
approaches the second particle and then decelerate. That means that
$\Delta_1 (x)$ is positive and so the common sign negative. Therefore,
attractive forces lead to a negative time delay in the center of momentum
frame, and clearly repulsive forces lead to a positive time delay. These
considerations assume that the two particles pass through each other, and
there is no reflection. However, when the masses of the two particles are
equal there is the possibility of occurring reflection.

In order to evaluate the time delay we choose the first particle to move
faster to the right than the second particle, i.e. $v_1 > v_2$, with $v_1 >
0$, and therefore from \rf{gammadef}  $|z_1| > |z_2|$ or
$\theta_1 > \theta_2$, where we have denoted $z_a = \eps_a
e^{\theta_a}$ with $\eps_a =\pm 1$.  Let us track the first soliton
in time \ct{fring}, i.e. hold $x-v_1 t$ fixed as time varies. One gets
\begin{equation}
x-v_2t= \underbrace{x-v_1 t}_{\mathrm{const}} + (v_1 - v_2)t.
\end{equation}
We then get from \rf{biggammadef} that, if $\eps_2 =1$,  $e^{\Gamma (z_2)}
\ra 0$ as $t\ra -\infty$, and $e^{\Gamma (z_2)} \ra \infty$ as $t\ra
\infty$ ($\Gamma$ stands for one of the three $\Gamma_i$ of
\rf{biggammadef}, corresponding to the second soliton). For the case
$\eps_2 = -1$ the limits get interchanged.

As we will see in section \ref{sec:twosol}, taking $\eps_2 =1$, all the
two-soliton solutions
become, in the limit $t\ra -\infty$, a one-soliton solution. Now, in the
limit
$t\ra \infty$, the two-solitons also become a one-soliton but with the
replacement
\be
e^{\Gamma (z_1)} \ra \( \frac{z_1-\omega z_2}{z_1+ \omega z_2}\)^n
\,e^{\Gamma (z_1)},
\lab{shiftg1a}
\ee
where $n$ is either $1$ or $2$, and $\omega =\pm 1$, depending
upon the type of two-soliton solution under consideration.
Therefore, the relevant effect of the scattering on the
solutions is a lateral displacement of the first soliton given by
\be
\gamma_1 \( x - v_1 t\) \ra  \gamma_1 \left[ x - v_1 t
+ \frac{1}{\gamma_1} \ln\( \frac{z_1-\omega z_2}{z_1+ \omega z_2}\)^n
\right].
\lab{shiftg1b}
\ee
If one takes $\eps_2 =-1$ instead, one observes that the direction  of the
arrow in \rf{shiftg1b} reverses.
Therefore, using \rf{latdisp} one sees that the lateral displacement for
the first soliton is given by
\be
\Delta_1 (x) = - \frac{\eps_2}{\gamma_1}
\ln\( \frac{z_1-\omega z_2}{z_1+\omega z_2}\)^n  =
- \frac{\eps_1\eps_2}{m \cosh \theta_1}
\ln\( \frac{ \rme^{\(\theta_1 - \theta_2\)/2} -
\eps_1\eps_2 \omega \rme^{-\(\theta_1 - \theta_2\)/2}}
{ \rme^{\(\theta_1 - \theta_2\)/2}+
\eps_1\eps_2 \omega \rme^{-\(\theta_1 - \theta_2\)/2}}\)^n,
\ee
where $m$ stands for one of the masses $m_i$, $i=1,2,3$, corresponding to
the first soliton. Since $\eps_1 \eps_2 \omega =\pm 1$, one observes that
$\Delta_1 (x)$ can in fact be written as
\be
\Delta_1 (x) = - \frac{\omega\, n}{m \cosh \theta_1}\,
\ln \left[ \tanh\( \frac{\theta_1-\theta_2}{2}\)\right].
\lab{deltax1}
\ee
Using \rf{timedelay} and \rf{deltax1}, one gets that the time delay is
given by (assuming $v_1 > v_2$, with $v_1>0$)
\be
\Delta_1 (t) = \frac{\omega\, n}{m \sinh \theta_1}\,
\ln \left[ \tanh\( \frac{\theta_1-\theta_2}{2}\) \right].
\lab{deltat}
\ee
Notice that we have taken $\theta_1>\theta_2$ and so the hyperbolic tangent
can vary from $0$ to $1$ and therefore its
logarithm  is always negative, and so from \rf{deltax1} one sees
that $\Delta_1(x)$ for $v_1 > v_2$, with $v_1>0$, has the same sign as
$\omega$. Therefore,  from  the above considerations we conclude
that the forces between the solitons are attractive for  $\omega =1$, and
repulsive for  $\omega =-1$. In section \ref{sec:twosol} we discuss the
details of each type of two-soliton solution.

\section{One-soliton solutions}
\label{sec:onesol}

As we have discussed in section \ref{sec:dressing} the one-soliton
solutions are obtained by taking the constant group element $\rho$,
parametrising the dressing transformations, as in \rf{ros}. Therefore,
we have a species of solitons for each positive root of
$\mathfrak{sl}_3(\bbC)$. We give below the results obtained by either
evaluating the matrix elements \rf{tauj} and \rf{taudef1}--\rf{taudef2}, or
by using the Hirota method (see program in appendix \ref{sec:app2}). As we
show below the one-soliton solution associated to a given root $\a_i$
excite only the tau-functions (or fields) associated to the same root.
Therefore, each species of solitons belongs to one of the three
$\mathfrak{sl}_2(\bbC)$ subalgebras associated to the positive roots of
$\mathfrak{sl}_3(\bbC)$.

We have checked that the three one-soliton  and six two-soliton solutions
constructed in this paper satisfy the equivalence between the vector and
topological currents given by \rf{equivcur}, or equivalently
\rf{equivcurtau1} and \rf{equivcurtau2}. Therefore, they are solutions of
the submodel defined by the constraints \rf{constraints}, and that presents
the confinement of the spinor charge inside the solitons.

\subsection{\mathversion{bold}One soliton of species $\a_1$}

In this case we have $\Sigma = \Sigma _1 = \exp \left[ \rme^{\G_1 (z)}
V_{\a_1}
(a_{\a_1}^\pm, z)\right]$. The only non-vanishing tau-functions are given
by
\br
& \tone = 1 - \frac{a^{-}_{\a_1}\,a^{+}_{\a_1}}{4} \,
{\rme^{2\,\Gamma_1(z)}}, \qquad \ttwo = 1 +
\frac{a^{-}_{\a_1}\,a^{+}_{\a_1}}{4} \, {\rme^{2\,\Gamma_1(z)}}, \qquad
\tthree = 1 - \frac{a^{-}_{\a_1}\,a^{+}_{\a_1}}{4} \,
{\rme^{2\,\Gamma_1(z)}}, \hspace{2em} \lab{onesol11} \\
& \tsixteen = a^{+}_{\a_1} \, z \, {\rme^{\Gamma_1(z)}}, \qquad \tseventeen
= a^{+}_{\a_1} \, z \, {\rme^{\Gamma_1(z)}}, \qquad \tseven =
a^{+}_{\a_1}\,{\rme^{\Gamma_1(z)}},  \\
& \tfour = a^{-}_{\a_1} \, {\rme^{\Gamma_1(z)}}, \qquad \tten = -
{\frac{a^{-}_{\a_1}\,}{z}} {\rme^{\Gamma_1(z)}}, \qquad \televen
=-{\frac{a^{-}_{\a_1}\,}{z}}{ \rme^{\Gamma_1(z)}}.
\lab{onesol12}
\er
As discussed below \rf{tautopchar}, in order to have $\vp_a$ real one needs
$|\tau_a| = |\tau_0|$, and consequently $a^{-}_{\a_1}\,a^{+}_{\a_1}$ has to
be pure imaginary. The topological current  \rf{tautopchar} is then
\be
Q^{\mathrm{top}} = -\frac{1}{2}\, \a_1 \( {\rm sign }\; z\)
\ee
and so it is (up to a sign) the fundamental weight of the
$\mathfrak{sl}_2(\bbC)$ subalgebra generated by $H_1$ and $E_{\pm \a_1}$.
Using \rf{solmass} and \rf{gammadef}, one obtains that the mass of the
soliton is $M^{\(\a_1\)}_{\mathrm{sol}} = 2 \kappa m_1$.

\subsection{\mathversion{bold}One soliton of species $\a_2$}

In this case we have $\Sigma = \Sigma _2 = \exp \left[ \rme^{\G_2\(z\)}
V_{\a_2}
(a_{\a_2}^{\pm}, z) \right]$. The non-vanishing tau-functions  are
given by
\br
& \tone = 1 - \frac{a^-_{\a_2} \, a^+_{\a_2}}{4} \,
{\rme^{2\,\Gamma_2(z)}},
\qquad \ttwo = 1 - \frac{a^-_{\a_2} \, a^+_{\a_2}}{4} \, {\rme^{2 \,
\Gamma_2(z)}}, \qquad \tthree = 1 + \frac{a^-_{\a_2} \, a^+_{\a_2}}{4} \,
{\rme^{2\,\Gamma_2(z)}}, \hspace{2em} \lab{onesol21} \\
& \teighteen = a^+_{\a_2} \, z \, \rme^{\Gamma_2(z)}, \qquad
\tnineteen = a^+_{\a_2} \, z \, \rme^{\Gamma_2(z)}, \qquad \teight =
a^+_{\a_2} \, {\rme^{\Gamma_2(z)}}, \\
& \tfive = a^-_{\a_2} \, \rme^{\Gamma_2(z)}, \qquad \ttwelve =
- \frac{a^-_{\a_2}}{z} \, \rme^{\Gamma_2(z)}, \qquad \tthirteen =
- \frac{a^-_{\a_2}}{z} \, \rme^{\Gamma_2(z)}.
\lab{onesol22}
\er
Similarly as above, in order to have $\vp_a$ real we need $a^-_{\a_2} \,
a^+_{\a_2}$ be pure imaginary. The topological current \rf{tautopchar} is
then
\be
Q^{\mathrm{top}} = -\frac{1}{2}\, \a_2 \( {\rm sign }\; z\)
\ee
and so it is (up to a sign) the fundamental weight of the
$\mathfrak{sl}_2(\bbC)$ subalgebra generated by $H_2$ and $E_{\pm \a_2}$.
The mass of the soliton is $M^{(\a_2)}_{\mathrm{sol}} = 2 \kappa m_2$.

\subsection{\mathversion{bold}One soliton of species $\a_3$}

In this case we have $\Sigma = \Sigma _3 = \exp \left[\rme^{\G_3(z)}
V_{\a_3} (a_{\a_3}^{\pm}, z)\right]$. The non-vanishing tau-functions  are
given by
\br
& \tone = 1 - \frac{a^-_{\a_3} \, a^+_{\a_3}}{4} \, \rme^{2 \,
\Gamma_3(z)},
\qquad \ttwo = 1 + \frac{a^-_{\a_3} \, a^+_{\a_3}}{4} \, \rme^{2 \,
\Gamma_3(z)}, \qquad \tthree = 1 + \frac{a^-_{\a_3} \, a^+_{\a_3}}{4} \,
\rme^{2 \, \Gamma_3(z)}, \hspace{2em} \lab{onesol31} \\
& \tsix = a^+_{\a_3} \, z \, \rme^{\Gamma_3(z)}, \qquad \tfourteen =
- a^+_{\a_3} \, \rme^{\Gamma_3(z)}, \qquad \tfifteen = - a^+_{\a_3} \,
\rme^{\Gamma_3(z)}, \\
& \ttwenty = a^-_{\a_3} \, \rme^{\Gamma_3(z)}, \ttwentyone = a^-_{\a_3} \,
\rme^{\Gamma_3(z)}, \qquad \tnine = - \frac{a^-_{\a_3}}{z} \,
\rme^{\Gamma_3(z)}.
\lab{onesol32}
\er
In order to have $\vp_a$ real we need $a^{-}_{\a_3}\,a^{+}_{\a_3}$ be pure
imaginary. The topological current \rf{tautopchar} is then
\be
Q^{\mathrm{top}} = -\frac{1}{2}\, \a_3 ( {\rm sign }\; z )
\ee
and so it is (up to a sign) the fundamental weight of the
$\mathfrak{sl}_2(\bbC)$ subalgebra generated by $H_1+H_2$ and $E_{\pm
\a_3}$. The mass of the soliton is $M^{(\a_3)}_{\mathrm{sol}} = 2 \kappa
m_3$.

\section{Two-soliton solutions}
\label{sec:twosol}

We now present the six two-soliton solutions corresponding to the pairs we
can form with the three species of solitons constructed in section
\ref{sec:onesol}. The solutions are constructed by performing the dressing
transformations with the constant group element $\rho$ leading to the
elements $\Sigma _{ij}$ given in \rf{gforsol2}. Instead of evaluating the
matrix elements in \rf{tauj} and \rf{taudef1}--\rf{taudef2} using the
homogeneous vertex operator realization of the three fundamental
representations of the $\cgh$, we calculated the solutions using the Hirota
method with ansatz \rf{gt}. The evaluation of the corresponding
coefficients was performed using the {\sf Mathematica} program described in
appendix~\ref{sec:app2}.

\subsection{\mathversion{bold}Two solitons of species $\a_1$/$\a_1$}

Here we give the two-soliton solution obtained by taking the group element
$\Sigma = \Sigma _{11} = \exp \left[\rme^{\G_1(z_1)} V_{\a_1}
(a_{\a_1}^{\pm}, z_1 )\right] \, \exp \left[\rme^{\G_1(z_2)} V_{\a_1}
(b_{\a_1}^{\pm}, z_2)\right]$. The non-vanishing tau-functions are
\br
\tone &=& 1 - \frac{a^-_{\a_1} a^+_{\a_1}}{4} \, \rme^{2\,\Gamma_1(z_1)} -
\frac{b^-_{\a_1} b^+_{\a_1}}{4} \, \rme^{2\,\Gamma_1(z_2)} -
\frac{(a^+_{\a_1} b^-_{\a_1} + a^-_{\a_1} b^+_{\a_1}) \, z_1 z_2}{(z_1 +
z_2)^2} \, \rme^{\Gamma_1(z_1) + \Gamma_1(z_2)} \nonu \\*
&& \hspace{8em} {} +  \frac{a^-_{\a_1} a^+_{\a_1} b^-_{\a_1} \, b^+_{\a_1}
\, (z_1 - z_2)^4}{16 \, (z_1 + z_2)^4} \, \rme^{2 \,\Gamma_1(z_1) +
2 \, \Gamma_1(z_2)}, \\
\ttwo &=&
1 + \frac{a^-_{\a_1} \, a^+_{\a_1}}{4} \, \rme^{2\,\Gamma_1(z_1)} +
\frac{b^-_{\a_1} \, b^+_{\a_1}}{4} \, \rme^{2\,\Gamma_1(z_2)}
+  \frac{(a^+_{\a_1} \, b^-_{\a_1} \, z_1^2 +       a^-_{\a_1} \, b^+_
{\a_1} \, z_2^2)}{( z_1 + z_2 )^2} \, \rme^{\Gamma_1(z_1) + \Gamma_1(z_2)}
\nonu \\*
&& \hspace{8em} {} + \frac{a^-_{\a_1} \, a^+_{\a_1} \, b^-_{\a_1} \,
b^+_{\a_1} \, (z_1 - z_2)^4}{16 (z_1 + z_2)^4} \, \rme^{2\,\Gamma_1(z_1) +
2\,\Gamma_1(z_2)}, \\
\tthree &=& 1 - \frac{a^-_{\a_1} \, a^+_{\a_1}}{4} \, \rme^{2 \,
\Gamma_1(z_1)} - \frac{b^-_{\a_1} \, b^+ _{\a_1}}{4} \, \rme^{2 \,
\Gamma_1(z_2)} - \frac{(a^+_{\a_1} \, b^-_{\a_1} +         a^-_{\a_1} \,
b^+_{\a_1} ) \, z_1 z_2}{(z_1 + z_2)^2} \, \rme^{\Gamma_1(z_1) +
\Gamma_1(z_2)} \nonu \\*
&& \hspace{8em} + \frac{a^-_{\a_1} \, a^+_{\a_1} \, b^-_{\a_1} \,
b^{+}_{\a_1} ( z_1 - z_2)^4}{16 \, (z_1 + z_2)^4} \, \rme^{2 \,
\Gamma_1(z_1) + 2 \, \Gamma_1(z_2)}, \\
\tfour &=& a^-_{\a_1} \, \rme^{\Gamma_1(z_1)} + b^-_{\a_1} \,
\rme^{\Gamma_1(z_2)} + \frac{a^-_{\a_1} \, a^+_{\a_1} \, b^-_{\a_1} \,
(z_1 - z_2)^2}{4 \, (z_1 + z_2)^2} \, \rme^{2 \,\Gamma_1(z_1) +
\Gamma_1(z_2)} \nonu \\*
&& \hspace{14em} {} + \frac{a^-_{\a_1} \, b^-_{\a_1} \, b^+_{\a_1}
\, (z_1 - z_2)^2}{4 \, (z_1 + z_2)^2} \, \rme^{\Gamma_1(z_1) + 2 \,
\Gamma_1(z_2)}, \\
\tseven &=& a^+_{\a_1} \, \rme^{\Gamma_1(z_1)} + b^+_{\a_1} \,
\rme^{\Gamma_1(z_2)} + \frac{a^-_{\a_1} \, a^+_{\a_1} \, b^+_{\a_1} \,
(z_1 - z_2)^2}{4 \, (z_1 + z_2)^2} \, \rme^{2 \, \Gamma_1(z_1) +
\Gamma_1(z_2)} \nonu \\*
&& \hspace{14em} {} + \frac{a^+_{\a_1} \, b^-_{\a_1} \, b^+_{\a_1} \,
(z_1 - z_2)^2}{4 \, (z_1 + z_2)^2} \, \rme^{\Gamma_1(z_1) + 2 \,
\Gamma_1(z_2)}, \\
\tten &=& - \frac{a^-_{\a_1}}{z_1} \, { \rme^{\Gamma_1(z_1)}} -
\frac{b^-_{\a_1}}{z_2} \, \rme^{\Gamma_1(z_2)} +   \frac{a^-_{\a_1} \,
b^-_{\a_1} \, b^+_{\a_1}\, (z_1 - z_2)^2}{4 \, z_1 \, (z_1 + z_2)^2} \,
\rme^{\Gamma_1(z_1) + 2 \, \Gamma_1(z_2)} \nonu \\*
&& \hspace{14em} {} + \frac{a^-_{\a_1} \, a^+_{\a_1} \, b^-_{\a_1} \,
(z_1 - z_2)^2}{4 \, z_2 \, (z_1 + z_2)^2} \, \rme^{2 \, \Gamma_1(z_1) +
\Gamma_1(z_2)}, \\
\televen &=& - \frac{a^-_{\a_1}}{z_1} \, \rme^{\Gamma_1(z_1)} -
\frac{b^-_{\a_1}}{z_2} \, \rme^{\Gamma_1(z_2)} +  \frac{a^-_{\a_1} \,
b^-_{\a_1} \, b^+_{\a_1} \, (z_1 - z_2)^2}{4 \, z_1 \, (z_1 + z_2)^2} \,
\rme^{\Gamma_1(z_1) + 2 \, \Gamma_1(z_2)} \nonu \\*
&& \hspace{14em} {} + \frac{a^-_{\a_1} \, a^+_{\a_1} \, b^-_{\a_1} \, (z_1
- z_2)^2}{4 \, z_2 \, (z_1 + z_2)^2} \, \rme^{2 \, \Gamma_1(z_1) +
\Gamma_1(z_2)}, \\
\tsixteen &=& a^+_{\a_1} \, z_1 \, \rme^{\Gamma_1(z_1)} + b^+_{\a_1} \, z_2
\, \rme^{\Gamma_1(z_2)} - \frac{a^+_{\a_1} \, b^-_{\a_1} \, b^+_{\a_1} \,
z_1\,(z_1 - z_2)^2}{4 \, (z_1 + z_2)^2} \, \rme^{\Gamma_1(z_1) + 2
\, \Gamma_1(z_2)} \nonu \\*
&& \hspace{14em} {} - \frac{a^-_{\a_1} \, a^+_{\a_1} \, b^+_{\a_1} \,
(z_1 - z_2)^2 \, z_2}{4 \, (z_1 + z_2)^2} \, \rme^{2 \, \Gamma_1(z_1) +
\Gamma_1(z_2)}, \\
\tseventeen &=& a^+_{\a_1} \, z_1 \, \rme^{\Gamma_1(z_1)} + b^+_{\a_1} \,
z_2 \rme^{\Gamma_1(z_2)} - \frac{a^+_{\a_1} \, b^-_{\a_1} \, b^+_{\a_1} \,
z_1 \, (z_1 - z_2)^2}{4 \, (z_1 + z_2)^2} \, \rme^{\Gamma_1(z_1) + 2 \,
\Gamma_1(z_2)} \nonu \\*
&& \hspace{14em} {} - \frac{a^-_{\a_1} \, a^+_{\a_1} \, b^+_{\a_1} \,
(z_1 - z_2)^2 \, z_2}{4 \, (z_1 + z_2)^2} \, \rme^{2 \, \Gamma_1(z_1) +
\Gamma_1(z_2)}.
\er
One can check that by taking the limit $\rme^{\G_1( z_2)} \ra 0$, such
solution becomes the one-soliton solution \rf{onesol11}--\rf{onesol12}.
Now, considering the ratios of tau-functions appearing in relations
\rf{phitau} and \rf{taucampos1}--\rf{taucampos2} one observes that in the
limit $\rme^{\G_1(
z_2)} \ra \infty$, such solution becomes the one-soliton solution
\rf{onesol11}--\rf{onesol12} again but with the replacement
\be
\rme^{\G_1( z_1)} \ra \( \frac{z_1-z_2}{z_1+z_2} \)^2 \, \rme^{\G_1( z_1)}
\ee
and with the ratio $\ttwo /\tone$ changing sign. Therefore, comparing with
\rf{shiftg1a}, we observe that the lateral displacement and time delay for
this two-soliton solution is given by \rf{deltax1} and \rf{deltat},
respectively, with $n = 2$ and $\omega = 1$. Consequently, the solitons of
species $\a_1$ experience an attractive force.

\subsection{\mathversion{bold}Two solitons of species $\a_2$/$\a_2$}

Here we give the two-soliton solutions obtained by taking the group element
$\Sigma = \Sigma _{22} = \exp \left[ \rme^{\G_2(z_1)} V_{\a_2}
(a_{\a_2}^{\pm}, z_1) \right] \, \exp \left[ \rme^{\G_2(z_2)} V_{\a_2}
(b_{\a_2}^{\pm}, z_2) \right]$. The non-vanishing tau-functions are
\br
\tone &=& 1 - \frac{a^-_{\a_2} \, a^+_{\a_2}}{4} \, \rme^{2 \,
\Gamma_2(z_1)} - \frac{b^-_{\a_2} \, b^+_{\a_2}}{4} \, \rme^{2 \,
\Gamma_2(z_2)} - \frac{(a^+_{\a_2} \, b^-_{\a_2} + a^-_{\a_2} \,
b^+_{\a_2}) \, z_1 \, z_2}{( z_1 + z_2)^2} \, \rme^{\Gamma_2(z_1) +
\Gamma_2(z_2)} \nonu \\*
&& \hspace{8em} {} + \frac{a^-_{\a_2} \, a^+_{\a_2} \, b^-_{\a_2} \,
b^+_{\a_2} \, (z_1 - z_2)^4}{16 \, (z_1 + z_2)^4} \, \rme^{2 \,
\Gamma_2(z_1) + 2 \, \Gamma_2(z_2)}, \\
\ttwo &=& 1 - \frac{a^-_{\a_2} \, a^+_{\a_2}}{4} \, \rme^{2\,\Gamma_2(z_1)}
- \frac{b^-_{\a_2} \, b^+_{\a_2}}{4} \, \rme^{2 \, \Gamma_2(z_2)} -
\frac{(a^+_{\a_2} \, b^-_{\a_2} + a^-_{\a_2} \, b^+_{\a_2}) \, z_1
\, z_2}{(z_1 + z_2)^2} \, \rme^{\Gamma_2(z_1) + \Gamma_2(z_2)} \nonu \\*
&& \hspace{8em} {} + \frac{a^-_{\a_2} \, a^+_{\a_2} \, b^-_{\a_2} \,
b^+_{\a_2} \, (z_1 - z_2)^4}{16 \, (z_1 + z_2)^4} \, \rme^{2 \,
\Gamma_2(z_1) + 2 \, \Gamma_2(z_2)}, \\
\tthree &=& 1 + \frac{a^-_{\a_2} \, a^+_{\a_2}}{4} \, \rme^{2 \,
\Gamma_2(z_1)} + \frac{b^-_{\a_2} \, b^+_{\a_2}}{4} \, \rme^{2 \,
\Gamma_2(z_2)} + \frac{(a^+_{\a_2} \, b^-_{\a_2} \, z_1^2 + a^-_{\a_2} \,
b^+_{\a_2} \, z_2^2)}{(z_1 + z_2)^2} \, \rme^{\Gamma_2(z_1) +
\Gamma_2(z_2)} \nonu \\*
&& \hspace{8em} {} + \frac{a^-_{\a_2} \, a^+_{\a_2} \, b^-_{\a_2} \,
b^{+}_{\a_2} \, (z_1 - z_2)^4}{16 \, (z_1 + z_2)^4} \, \rme^{2 \,
\Gamma_2(z_1) + 2 \, \Gamma_2(z_2)}, \\
\tfive &=& a^-_{\a_2} \, \rme^{\Gamma_2(z_1)} + b^-_{\a_2} \,
\rme^{\Gamma_2(z_2)} + \frac{a^-_{\a_2} \, a^+_{\a_2} \, b^{-}_{\a_2} \,
(z_1 - z_2)^2}{4 \, (z_1 + z_2)^2} \, \rme^{2 \, \Gamma_2(z_1) +
\Gamma_2(z_2)} \nonu \\*
&& \hspace{14em} {} + \frac{a^-_{\a_2} \, b^-_{\a_2} \, b^+_{\a_2} \, (z_1
- z_2)^2}{4 \, (z_1 + z_2)^2} \, \rme^{\Gamma_2(z_1) + 2 \, \Gamma_2(z_2)},
\\
\teight &=& a^+_{\a_2} \, \rme^{\Gamma_2(z_1)} + b^+_{\a_2} \,
\rme^{\Gamma_2(z_2)} + \frac{a^-_{\a_2} \, a^+_{\a_2} \, b^+_{\a_2} \, (z_1
- z_2)^2}{4 \, (z_1 + z_2)^2} \, \rme^{2 \, \Gamma_2(z_1) + \Gamma_2(z_2)}
\nonu \\*
&& \hspace{14em} + \frac{a^+_{\a_2} \, b^-_{\a_2} \, b^+_{\a_2} \, (z_1 -
z_2)^2}{4 \, (z_1 + z_2)^2} \, \rme^{\Gamma_2(z_1) + 2 \, \Gamma_2(z_2)},
\\
\ttwelve &=& - \frac{a^-_{\a_2}}{z_1} \, \rme^{\Gamma_2(z_1)} -
\frac{b^-_{\a_2}}{z_2} \, \rme^{\Gamma_2(z_2)} + \frac{a^-_{\a_2} \,
b^-_{\a_2} \, b^+_{\a_2} \, (z_1 - z_2)^2}{4 \, z_1 \, (z_1 + z_2)^2} \,
\rme^{\Gamma_2(z_1) + 2 \, \Gamma_2(z_2)} \nonu \\*
&& \hspace{14em} {} + \frac{a^-_{\a_2} \, a^+_{\a_2} \, b^-_{\a_2} \, (z_1
- z_2)^2}{4 \, z_2 \, (z_1 + z_2)^2} \, \rme^{2 \, \Gamma_2(z_1) +
\Gamma_2(z_2)}, \\
\tthirteen &=& -\frac{a^-_{\a_2}}{z_1} \, \rme^{\Gamma_2(z_1)} -
\frac{b^-_{\a_2}}{z_2} \, \rme^{\Gamma_2(z_2)}
+ \frac{a^-_{\a_2} \, b^-_{\a_2} \, b^+_{\a_2} \, (z_1 - z_2)^2}{4 \, z_1
\, (z_1 + z_2)^2} \, \rme^{\Gamma_2(z_1) + 2 \, \Gamma_2(z_2)} \nonu \\*
&& \hspace{14em} {} + \frac{a^-_{\a_2} \, a^+_{\a_2} \, b^-_{\a_2} \, (z_1
- z_2)^2}{4 \, z_2 \, (z_1 + z_2)^2} \, \rme^{2\,\Gamma_2(z_1) +
\Gamma_2(z_2)}, \\
\teighteen &=& a^+_{\a_2} \, z_1 \, \rme^{\Gamma_2(z_1)} + b^+_{\a_2} \,
z_2 \, \rme^{\Gamma_2(z_2)} - \frac{a^+_{\a_2} \, b^-_{\a_2} \,
b^+_{\a_2} \, z_1 \, (z_1 - z_2)^2}{4 \, (z_1 + z_2)^2} \,
\rme^{\Gamma_2(z_1) + 2 \, \Gamma_2(z_2)} \nonu \\*
&& \hspace{14em} {} - \frac{a^-_{\a_2} \, a^+_{\a_2} \, b^{+}_{\a_2} \,
(z_1 - z_2)^2 \, z_2}{4 \, (z_1 + z_2)^2} \, \rme^{2 \, \Gamma_2(z_1) +
\Gamma_2(z_2)}, \\
\tnineteen &=& a^+_{\a_2} \, z_1 \, \rme^{\Gamma_2(z_1)} + b^+_{\a_2} \,
z_2 \, \rme^{\Gamma_2(z_2)} - \frac{a^+_{\a_2} \, b^-_{\a_2} \, b^+_{\a_2}
\, z_1 \, (z_1 - z_2)^2}{4 \, (z_1 + z_2)^2} \, \rme^{\Gamma_2(z_1) + 2 \,
\Gamma_2(z_2)} \nonu \\*
&& \hspace{14em} {} - \frac{a^-_{\a_2} \, a^+_{\a_2} \, b^+_{\a_2} \, (z_1
- z_2)^2 \, z_2}{4 \, (z_1 + z_2)^2} \, \rme^{2\,\Gamma_2(z_1) +
\Gamma_2(z_2)}.
\er
One can check that by taking the limit $\rme^{\G_2(z_2)} \ra 0$, such
solution becomes the one-soliton solution \rf{onesol21}--\rf{onesol22}.
Now, considering the ratios of tau-functions appearing in the relations
\rf{phitau} and \rf{taucampos1}--\rf{taucampos2} one observes that in the
limit
$\rme^{\G_2(z_2)} \ra \infty$, such solution becomes the one-soliton
solution \rf{onesol21}--\rf{onesol22} again, but with the replacement
\be
\rme^{\G_2(z_1)} \ra \( \frac{z_1-z_2}{z_1+z_2}\)^2 \, \rme^{\G_2\( z_1\)}
\ee
and with the ratio $\tthree /\tone$ changing sign. Therefore, comparing
with \rf{shiftg1a}, we observe that the lateral displacement and time delay
for this two-soliton solution is given by \rf{deltax1} and \rf{deltat},
respectively, with $n = 2$ and $\omega = 1$. Consequently, the solitons of
species $\a_2$ experience an attractive force.

\subsection{\mathversion{bold}Two solitons of species $\a_3$/$\a_3$}

Here we give the two-soliton solutions obtained by taking the group element
$\Sigma = \Sigma _{33} = \exp \left[ \rme^{\G_3(z_1)} V_{\a_3}
(a_{\a_3}^{\pm}, z_1) \right] \, \exp \left[\rme^{\G_3(z_2)} V_{\a_3}
(b_{\a_3}^{\pm}, z_2 ) \right]$. The non-vanishing tau-functions are
\br
\tone &=& 1 - \frac{a^-_{\a_3} \, a^+_{\a_3}}{4} \, \rme^{2 \,
\Gamma_3(z_1)} - \frac{b^-_{\a_3} \, b^+_{\a_3}}{4} \, \rme^{2 \,
\Gamma_3(z_2)} - \frac{(a^+_{\a_3} \, b^-_{\a_3} + a^-_{\a_3} \,
b^+_{\a_3}) \, z_1 \, z_2}{(z_1 + z_2)^2} \, \rme^{\Gamma_3(z_1) +
\Gamma_3(z_2)} \nonu \\*
&& \hspace{8em} {} + \frac{a^-_{\a_3} \, a^+_{\a_3} \, b^-_{\a_3} \,
b^+_{\a_3} \, (z_1 - z_2)^4}{16 \, (z_1 + z_2)^4} \, \rme^{2 \,
\Gamma_3(z_1) + 2 \, \Gamma_3(z_2)}, \\
\ttwo &=& 1 + \frac{a^-_{\a_3} \, a^+_{\a_3}}{4} \, \rme^{2 \,
\Gamma_3(z_1)} + \frac{b^-_{\a_3} \, b^+_{\a_3}}{4} \,
\rme^{2\,\Gamma_3(z_2)} + \frac{(a^+_{\a_3} \, b^-_{\a_3} \, z_1^2 +
a^-_{\a_3} \, b^+_{\a_3} \, z_2^2)}{(z_1 + z_2)^2} \, \rme^{\Gamma_3(z_1) +
\Gamma_3(z_2)} \nonu \\*
&& \hspace{8em} {} + \frac{a^-_{\a_3} \, a^+_{\a_3} \, b^{-}_{\a_3} \,
b^{+}_{\a_3} \, (z_1 - z_2)^4}{16 \, (z_1 + z_2)^4} \, \rme^{2 \,
\Gamma_3(z_1) + 2 \, \Gamma_3(z_2)}, \\
\tthree &=& 1 + \frac{a^-_{\a_3} \, a^+_{\a_3}}{4} \, \rme^{2 \,
\Gamma_3(z_1)} + \frac{b^-_{\a_3} \, b^+_{\a_3}}{4} \, \rme^{2 \,
\Gamma_3(z_2)} + \frac{(a^+_{\a_3} \, b^-_{\a_3} \, z_1^2 + a^-_{\a_3} \,
b^+_{\a_3} \, z_2^2)}{(z_1 + z_2)^2} \, \rme^{\Gamma_3(z_1) +
\Gamma_3(z_2)} \nonu \\*
&& \hspace{8em} {} + \frac{a^-_{\a_3} \, a^+_{\a_3} \, b^-_{\a_3} \,
b^+_{\a_3} (z_1 - z_2)^4}{16 \, (z_1 + z_2)^4} \, \rme^{2 \, \Gamma_3(z_1)
+ 2 \, \Gamma_3(z_2)}, \\
\tsix &=& a^+_{\a_3} \, z_1 \, \rme^{\Gamma_3(z_1)} z_1 + b^+_{\a_3} \, z_2
\, \rme^{\Gamma_3(z_2)} - \frac{a^+_{\a_3} \, b^-_{\a_3} \, b^+_{\a_3}
\,z_1 \, (z_1 - z_2)^2}{4 \, (z_1 + z_2)^2} \, \rme^{\Gamma_3(z_1) + 2 \,
\Gamma_3(z_2)} \nonu \\*
&& \hspace{14em} {} - \frac{a^-_{\a_3} \, a^+_{\a_3} \, b^+_{\a_3} \, (z_1
- z_2)^2 \, z_2}{4 \, (z_1 + z_2)^2} \, \rme^{2\,\Gamma_3(z_1) +
\Gamma_3(z_2)}, \\
\tnine &=& - \frac{a^-_{\a_3}}{z_1} \, \rme^{\Gamma_3(z_1)} -
\frac{b^-_{\a_3}}{z_2} \, \rme^{\Gamma_3(z_2)} + \frac{a^-_{\a_3} \,
b^{-}_{\a_3} \, b^+_{\a_3} \, (z_1 - z_2)^2}{4 \, z_1\, (z_1 + z_2)^2} \,
\rme^{\Gamma_3(z_1) + 2 \, \Gamma_3(z_2)} \nonu \\*
&& \hspace{14em} {} + \frac{a^-_{\a_3} \, a^+_{\a_3} \, b^-_{\a_3} \, (z_1
- z_2)^2}{4 \, z_2 \, (z_1 + z_2)^2} \, \rme^{2 \,\Gamma_3(z_1) +
\Gamma_3(z_2)}, \\
\tfourteen &=& - a^+_{\a_3} \, \rme^{\Gamma_3(z_1)} - b^+_{\a_3} \,
\rme^{\Gamma_3(z_2)} - \frac{a^-_{\a_3} \, a^+_{\a_3} \, b^+_{\a_3} \, (z_1
- z_2)^2}{4 \, (z_1 + z_2)^2} \, \rme^{2\,\Gamma_3(z_1) + \Gamma_3(z_2)}
\nonu \\*
&& \hspace{14em} {} - \frac{a^+_{\a_3} \, b^-_{\a_3} \, b^+_{\a_3} \, (z_1
- z_2)^2}{4 \, (z_1 + z_2)^2} \, \rme^{\Gamma_3(z_1) + 2 \, \Gamma_3(z_2)},
\\
\tfifteen &=& - a^+_{\a_3} \, \rme^{\Gamma_3(z_1)} - b^+_{\a_3} \,
\rme^{\Gamma_3(z_2)} - \frac{a^-_{\a_3} \, a^+_{\a_3} \, b^+_{\a_3} \, (z_1
- z_2)^2}{4 \, (z_1 + z_2)^2} \, \rme^{2 \, \Gamma_3(z_1) + \Gamma_3(z_2)}
\nonu \\*
&& \hspace{14em} {}- \frac{a^+_{\a_3} \, b^-_{\a_3} \, b^+_{\a_3} \, (z_1 -
z_2)^2}{4 \, (z_1 + z_2)^2} \, \rme^{\Gamma_3(z_1) + 2 \, \Gamma_3(z_2)},
\\
\ttwenty &=& a^-_{\a_3} \, \rme^{\Gamma_3(z_1)} + b^-_{\a_3} \,
\rme^{\Gamma_3(z_2)} + \frac{a^-_{\a_3} \, a^+_{\a_3} \, b^-_{\a_3} \, (z_1
- z_2)^2}{4 \, (z_1 + z_2)^2} \, \rme^{2 \, \Gamma_3(z_1) + \Gamma_3(z_2)}
\nonu \\*
&& \hspace{14em} {} + \frac{a^-_{\a_3} \, b^-_{\a_3} \, b^+_{\a_3} \, (z_1
- z_2)^2}{4 \, (z_1 + z_2)^2} \, \rme^{\Gamma_3(z_1) + 2 \, \Gamma_3(z_2)},
\\
\ttwentyone &=& a^-_{\a_3} \, \rme^{\Gamma_3(z_1)} +  b^-_{\a_3} \,
\rme^{\Gamma_3(z_2)} + \frac{a^-_{\a_3} \, a^+_{\a_3} \, b^-_{\a_3} \, (z_1
- z_2)^2}{4 \, (z_1 + z_2)^2} \, \rme^{2 \, \Gamma_3(z_1) + \Gamma_3(z_2)}
\nonu \\*
&& \hspace{14em} {}+ \frac{a^-_{\a_3} \, b^-_{\a_3} \, b^+_{\a_3} \, (z_1 -
z_2)^2}{4 \, (z_1 + z_2)^2} \, \rme^{\Gamma_3(z_1) + 2 \, \Gamma_3(z_2)}.
\er
One can check that by taking the limit $\rme^{\G_3(z_2)} \ra 0$, such
solution become the one-soliton solution \rf{onesol31}--\rf{onesol32}. Now,
considering the ratios of tau-functions appearing in the relations
\rf{phitau} and \rf{taucampos1}--\rf{taucampos2} one observes that in the
limit
$\rme^{\G_3(z_2)} \ra \infty$, such solution becomes the one-soliton
solution \rf{onesol31}--\rf{onesol32} again, but with the replacement
\be
\rme^{\G_3(z_1)} \ra \( \frac{z_1-z_2}{z_1+z_2}\)^2 \, \rme^{\G_3(z_1)}
\ee
and with the ratios $\ttwo /\tone$ and $\tthree /\tone$ changing signs.
Therefore, comparing with \rf{shiftg1a}, we observe that the lateral
displacement and time delay for this two-soliton solution is given by
\rf{deltax1} and \rf{deltat}, respectively, with $n = 2$ and $\omega = 1$.
Consequently, the solitons of species $\a_3$ experience an attractive
force.

\subsection{\mathversion{bold}Two solitons of species $\a_1$/$\a_2$}

Here we give the two-soliton solutions obtained by taking the group element
$\Sigma = \Sigma _{12} = \exp \left[ \rme^{\G_1(z_1)} V_{\a_1}
(a_{\a_1}^{\pm}, z_1) \right] \, \exp \left[ \rme^{\G_2(z_2)} V_{\a_2}
(a_{\a_2}^{\pm}, z_2) \right]$. The non-vanishing tau-functions are
\br
\tone &=& 1 - \frac{a^-_{\a_1} \, a^+_{\a_1}}{4} \, \rme^{2 \,
\Gamma_1(z_1)} - \frac{a^-_{\a_2} \, a^+_{\a_2}}{4} \, \rme^{2 \,
\Gamma_2(z_2)} \nonu \\*
&& \hspace{6em} {} + \frac{a^-_{\a_1} \, a^-_{\a_2} \, a^+_{\a_1} \,
a^{+}_{\a_2} \, (z_1 + z_2)^2}{16 \, (z_1 - z_2)^2} \, \rme^{2 \,
\Gamma_1(z_1) + 2 \, \Gamma_2(z_2)}, \\
\ttwo &=& 1 + \frac{a^-_{\a_1} \, a^+_{\a_1}}{4} \, \rme^{2 \,
\Gamma_1(z_1)} - \frac{a^-_{\a_2} \, a^+_{\a_2}}{4} \, \rme^{2 \,
\Gamma_2(z_2)} \nonu \\*
&& \hspace{6em} - \frac{a^-_{\a_1} \, a^-_{\a_2} \, a^+_{\a_1} \,
a^+_{\a_2} \, (z_1 + z_2)^2}{16 \, (z_1 - z_2)^2} \, \rme^{2 \,
\Gamma_1(z_1) + 2 \, \Gamma_2(z_2)}, \\
\tthree &=& 1 - \frac{a^-_{\a_1} \, a^+_{\a_1}}{4} \, \rme^{2 \,
\Gamma_1(z_1)} + \frac{a^-_{\a_2} \, a^+_{\a_2}}{4} \, \rme^{2 \,
\Gamma_2(z_2)} \nonu \\*
&& \hspace{6em} - \frac{a^-_{\a_1} \, a^-_{\a_2} \, a^+_{\a_1} \,
a^+_{\a_2} \, (z_1 + z_2)^2}{16 \, (z_1 - z_2)^2} \, \rme^{2 \,
\Gamma_1(z_1) + 2 \, \Gamma_2(z_2)}, \\
\tfour &=& a^-_{\a_1} \,  \rme^{\Gamma_1(z_1)} - \frac{a^-_{\a_1} \,
a^-_{\a_2} \, a^+_{\a_2} \, ( z_1 + z_2)}{4 \, (z_1 - z_2)} \,
\rme^{\Gamma_1(z_1) + 2 \, \Gamma_2(z_2)}, \\
\tfive &=& a^-_{\a_2} \, \rme^{\Gamma_2(z_2)} + \frac{a^-_{\a_1} \,
a^-_{\a_2} \, a^+_{\a_1} \, (z_1 + z_2)}{4 \, (z_1 - z_2)} \, \rme^{2 \,
\Gamma_1(z_1) + \Gamma_2(z_2)}, \\
\tsix &=& \frac{a^+_{\a_1} \, a^+_{\a_2}\, z_1 \, z_2}{z_1 - z_2} \,
\rme^{\Gamma_1(z_1) + \Gamma_2(z_2)}, \\
\tseven &=& a^+_{\a_1} \, \rme^{\Gamma_1(z_1)} + \frac{a^-_{\a_2} \,
a^+_{\a_1} \, a^+_{\a_2} \, (z_1 + z_2)}{4 \, (z_1 - z_2)} \,
\rme^{\Gamma_1(z_1) + 2 \, \Gamma_2(z_2)}, \\
\teight &=& a^+_{\a_2} \, \rme^{\Gamma_2(z_2)} - \frac{a^-_{\a_1} \,
a^+_{\a_1} \, a^+_{\a_2} \, (z_1 + z_2)}{4 \, (z_1 - z_2)} \, \rme^{2 \,
\Gamma_1(z_1) + \Gamma_2(z_2)}, \\
\tnine &=& \frac{a^-_{\a_1} \, a^-_{\a_2}}{z_1 - z_2} \,
\rme^{\Gamma_1(z_1) + \Gamma_2(z_2)}, \\
\tten &=& -\frac{a^-_{\a_1}}{z_1} \, \rme^{\Gamma_1(z_1)} +
\frac{a^-_{\a_1} \, a^-_{\a_2} \, a^+_{\a_2} \, (z_1 + z_2)}{4 \, z_1 \,
(z_1 - z_2)} \, \rme^{\Gamma_1(z_1) + 2 \, \Gamma_2(z_2)}, \\
\televen &=& -\frac{a^-_{\a_1}}{z_1} \, \rme^{\Gamma_1(z_1)} -
\frac{a^-_{\a_1} \, a^-_{\a_2} \, a^+_{\a_2} \, (z_1 + z_2)}{4 \, z_1 \,
(z_1 - z_2)} \, \rme^{\Gamma_1(z_1) + 2 \, \Gamma_2(z_2)}, \\
\ttwelve &=& -\frac{a^-_{\a_2}}{z_2} \, \rme^{\Gamma_2(z_2)} -
\frac{a^-_{\a_1} \, a^-_{\a_2} \, a^+_{\a_1} \, (z_1 + z_2)}{4 \, (z_1 -
z_2) \, z_2} \, \rme^{2 \, \Gamma_1(z_1) + \Gamma_2(z_2)}, \\
\tthirteen &=& -\frac{a^-_{\a_2}}{z_2} \, \rme^{\Gamma_2(z_2)} +
\frac{a^-_{\a_1} \, a^-_{\a_2} \, a^+_{\a_1} \, (z_1 + z_2)}{4 \, (z_1 -
z_2) \, z_2} \, \rme^{2 \, \Gamma_1(z_1) + \Gamma_2(z_2)}, \\
\tfourteen &=& - \frac{a^+_{\a_1} \, a^+_{\a_2} \, z_1}{z_1 - z_2} \,
\rme^{\Gamma_1(z_1) + \Gamma_2(z_2)}, \\
\tfifteen &=& -\frac{a^+_{\a_1} \, a^+_{\a_2} \, z_2}{z_1 - z_2} \,
\rme^{\Gamma_1(z_1) + \Gamma_2(z_2)}, \\
\tsixteen &=& a^+_{\a_1} \, z_1 \, \rme^{\Gamma_1(z_1)} - \frac{a^-_{\a_2}
\, a^+_{\a_1} \, a^+_{\a_2} \, z_1 \, (z_1 + z_2)}{4 \, (z_1 - z_2)} \,
\rme^{\Gamma_1(z_1) + 2 \, \Gamma_2(z_2)}, \\
\tseventeen &=& a^+_{\a_1} \, z_1 \, \rme^{\Gamma_1(z_1)} +
\frac{a^-_{\a_2} \, a^+_{\a_1} \, a^+_{\a_2} \, z_1 \, (z_1 + z_2)}{4 \,
(z_1 - z_2)} \, \rme^{\Gamma_1(z_1) + 2 \, \Gamma_2(z_2)}, \\
\teighteen &=& a^+_{\a_2} \, z_2 \, \rme^{\Gamma_2(z_2)} + \frac{a^-_{\a_1}
\, a^+_{\a_1} \, a^+_{\a_2}\,z_2 \, (z_1 + z_2)}{4 \, (z_1 - z_2)} \,
\rme^{2 \, \Gamma_1(z_1) + \Gamma_2(z_2)}, \\
\tnineteen &=& a^+_{\a_2} \, z_2 \, \rme^{\Gamma_2(z_2)} - \frac{a^-_{\a_1}
\, a^+_{\a_1} \, a^+_{\a_2} \,z_2 \, (z_1 + z_2)}{4 \, (z_1 - z_2)} \,
\rme^{2 \, \Gamma_1(z_1) + \Gamma_2(z_2)}, \\
\ttwenty &=& - \frac{a^-_{\a_1} \, a^-_{\a_2} \, z_2}{z_1 - z_2} \,
\rme^{\Gamma_1(z_1) + \Gamma_2(z_2)}, \\
\ttwentyone &=& -\frac{a^-_{\a_1} \, a^-_{\a_2} \, z_1}{z_1 - z_2} \,
\rme^{\Gamma_1(z_1) + \Gamma_2(z_2)}.
\er
One can check that by taking the limit $\rme^{\G_2(z_2)} \ra 0$, such
solution becomes the one-soliton solution \rf{onesol11}--\rf{onesol12}.
Now, considering the ratios of tau-functions appearing in the relations
\rf{phitau} and \rf{taucampos1}--\rf{taucampos2} one observes that in the
limit
$\rme^{\G_2(z_2)} \ra \infty$, such solution becomes the one-soliton
solution \rf{onesol11}--\rf{onesol12} again, but with the replacement
\be
\rme^{\G_1(z_1)} \ra \( \frac{z_1+z_2}{z_1-z_2}\) \, \rme^{\G_1(z_1)}
\ee
and with the ratios $\tthree /\tone$, $\tseven /\ttwo$, $\tten / \tthree$,
and $\televen /\tone$  changing signs. If we track the soliton $\a_2$
instead of $\a_1$, we get the same results (interchanging $\G_1
\leftrightarrow \G_2$), but the ratios of tau-functions changing signs are
$\ttwo /\tone$, $\tfive / \tthree$, $\teighteen /\tone$, and $\tnineteen
/\ttwo$. Therefore, comparing with \rf{shiftg1a}, we observe that the
lateral displacement and time delay for this two-soliton solution is given
by \rf{deltax1} and \rf{deltat}, respectively, with $n = 1$ and $\omega =
-1$. Consequently, the solitons of species $\a_1$ and $\a_2$ experience a
repulsive force.

\subsection{\mathversion{bold}Two solitons of species $\a_1$/$\a_3$}

Here we give the two-soliton solutions obtained by taking the group element
$\Sigma = \Sigma _{13} = \exp \left[ \rme^{\G_1(z_1)} V_{\a_1}
(a_{\a_1}^{\pm}, z_1) \right] \, \exp \left[ \rme^{\G_3(z_2)} V_{\a_3}
(a_{\a_3}^{\pm}, z_2) \right]$. The non-vanishing tau-functions are
\br
\tone &=& 1 - \frac{a^-_{\a_1} \, a^+_{\a_1}}{4} \, \rme^{2 \,
\Gamma_1(z_1)} - \frac{a^-_{\a_3} \, a^+_{\a_3}}{4} \,  \rme^{2 \,
\Gamma_3(z_2)} \nonu \\*
&& \hspace{6em} + \frac{a^-_{\a_1} \, a^-_{\a_3} \, a^+_{\a_1} \,
a^{+}_{\a_3} \, (z_1 - z_2)^2}{16 \, (z_1 + z_2)^2} \, \rme^{2 \,
\Gamma_1(z_1) + 2 \, \Gamma_3(z_2)}, \\
\ttwo &=& 1 + \frac{a^-_{\a_1} \, a^+_{\a_1}}{4} \, \rme^{2 \,
\Gamma_1(z_1)} + \frac{a^-_{\a_3} \, a^+_{\a_3}}{4} \,
\rme^{2\,\Gamma_3(z_2)} \nonu \\*
&& \hspace{6em} + \frac{a^-_{\a_1} \, a^-_{\a_3} \, a^+_{\a_1} \,
a^+_{\a_3} \, (z_1 - z_2)^2}{16 \, (z_1 + z_2)^2} \, \rme^{2 \,
\Gamma_1(z_1) + 2 \, \Gamma_3(z_2)}, \\
\tthree &=& 1 - \frac{a^-_{\a_1} \, a^+_{\a_1}}{4} \, \rme^{2 \,
\Gamma_1(z_1)} + \frac{a^-_{\a_3} \, a^+_{\a_3}}{4} \,  \rme^{2 \,
\Gamma_3(z_2)} \nonu \\*
&& \hspace{6em} - \frac{a^-_{\a_1} \, a^-_{\a_3} \, a^+_{\a_1} \,
a^+_{\a_3} \, (z_1 - z_2)^2}{16 \, (z_1 + z_2)^2} \, \rme^{2 \,
\Gamma_1(z_1) + 2 \, \Gamma_3(z_2)}, \\
\tfour &=& a^-_{\a_1} \, \rme^{\Gamma_1(z_1)} + \frac{a^-_{\a_1} \,
a^-_{\a_3} \, a^+_{\a_3} \, (z_1 - z_2)}{4 \, (z_1 + z_2)} \,
\rme^{\Gamma_1(z_1) + 2 \, \Gamma_3(z_2)}, \\
\tfive &=& -\frac{a^-_{\a_3} \, a^+_{\a_1} \, z_1}{z_1 + z_2} \,
\rme^{\Gamma_1(z_1) + \Gamma_3(z_2)}, \\
\tsix &=& a^+_{\a_3} \, z_2 \, \rme^{\Gamma_3(z_2)} + \frac{a^-_{\a_1} \,
a^+_{\a_1} \, a^+_{\a_3} \, (z_1 - z_2) \, z_2}{4 \, (z_1 + z_2)} \,
\rme^{2 \, \Gamma_1(z_1) + \Gamma_3(z_2)}, \\
\tseven &=& a^+_{\a_1} \, \rme^{\Gamma_1(z_1)} - \frac{a^-_{\a_3} \,
a^+_{\a_1} \, a^+_{\a_3} \, (z_1 - z_2)}{4 \, (z_1 + z_2)} \,
\rme^{\Gamma_1(z_1) + 2 \, \Gamma_3(z_2)}, \\
\teight &=& -\frac{a^-_{\a_1} \, a^+_{\a_3} \, z_2}{z_1 + z_2} \,
\rme^{\Gamma_1(z_1) + \Gamma_3(z_2)}, \\
\tnine &=& -\frac{a^-_{\a_3}}{z_2} \, \rme^{\Gamma_3(z_2)} +
\frac{a^-_{\a_1} \, a^-_{\a_3} \, a^+_{\a_1} \, (z_1 - z_2)}{4 \, z_2 \,
(z_1 + z_2)} \, \rme^{2 \, \Gamma_1(z_1) + \Gamma_3(z_2)}, \\
\tten &=& -\frac{a^-_{\a_1}}{z_1} \, \rme^{\Gamma_1(z_1)} +
\frac{a^-_{\a_1} \, a^-_{\a_3} \, a^+_{\a_3} \, (z_1 - z_2)}{4 \, z_1 \,
(z_1 + z_2)} \, \rme^{\Gamma_1(z_1) + 2 \, \Gamma_3(z_2)}, \\
\televen &=& -\frac{a^-_{\a_1}}{z_1} \, \rme^{\Gamma_1(z_1)} -
\frac{a^-_{\a_1} \, a^-_{\a_3} \, a^+_{\a_3} \, (z_1 - z_2)}{4 \, z_1 \,
(z_1 + z_2)} \, \rme^{\Gamma_1(z_1) + 2 \, \Gamma_3(z_2)}, \\
\ttwelve &=& \frac{a^-_{\a_3} \, a^+_{\a_1} \, z_1}{z_2 \, (z_1 + z_2)} \,
\rme^{\Gamma_1(z_1) + \Gamma_3(z_2)}, \\
\tthirteen &=& -\frac{a^-_{\a_3} \, a^+_{\a_1}}{z_1 + z_2} \,
\rme^{\Gamma_1(z_1) + \Gamma_3(z_2)}, \\
\tfourteen &=& - a^+_{\a_3} \, \rme^{\Gamma_3(z_2)} - \frac{a^-_{\a_1} \,
a^+_{\a_1} \, a^+_{\a_3} \, (z_1 - z_2)}{4 \, (z_1 + z_2)} \, \rme^{2 \,
\Gamma_1(z_1) + \Gamma_3(z_2)}, \\
\tfifteen &=& -a^+_{\a_3} \, \rme^{\Gamma_3(z_2)} + \frac{a^-_{\a_1} \,
a^+_{\a_1} \, a^+_{\a_3} \, (z_1 - z_2)}{4 \, ( z_1 + z_2)} \, \rme^{2 \,
\Gamma_1(z_1) + \Gamma_3(z_2)}, \\
\tsixteen &=& a^+_{\a_1} \, z_1 \, \rme^{\Gamma_1(z_1)} -
\frac{a^-_{\a_3} \, a^+_{\a_1} \, a^+_{\a_3} \, z_1 \, (z_1 - z_2)}{4 \,
(z_1 + z_2)} \, \rme^{\Gamma_1(z_1) + 2 \, \Gamma_3(z_2)}, \\
\tseventeen &=& a^+_{\a_1} \, z_1 \, \rme^{\Gamma_1(z_1)} +
\frac{a^-_{\a_3} \, a^+_{\a_1} \, a^+_{\a_3} \, z_1 \, (z_1 - z_2)}{4 \,
(z_1 + z_2)} \, \rme^{\Gamma_1(z_1) + 2 \, \Gamma_3(z_2)}, \\
\teighteen &=& \frac{a^-_{\a_1} \, a^+_{\a_3} \,z_1 \, z_2}{z_1 + z_2} \,
\rme^{\Gamma_1(z_1) + \Gamma_3(z_2)}, \\
\tnineteen &=& -\frac{a^-_{\a_1} \, a^{+}_{\a_3} \, z_2^2}{z_1 + z_2} \,
\rme^{\Gamma_1(z_1) + \Gamma_3(z_2)}, \\
\ttwenty &=& a^-_{\a_3} \, \rme^{\Gamma_3(z_2)} - \frac{a^-_{\a_1} \,
a^-_{\a_3} \, a^+_{\a_1} \, (z_1 - z_2)}{4 \, (z_1 + z_2)} \, \rme^{2 \,
\Gamma_1(z_1) + \Gamma_3(z_2)}, \\
\ttwentyone &=& a^-_{\a_3} \, \rme^{\Gamma_3(z_2)} + \frac{a^-_{\a_1} \,
a^-_{\a_3} \, a^+_{\a_1} \, (z_1 - z_2)}{4 \, (z_1 + z_2)} \, \rme^{2 \,
\Gamma_1(z_1) + \Gamma_3(z_2)}.
\er
One can check that by taking the limit $\rme^{\G_3(z_2)} \ra 0$, such
solution become the one-soliton solution \rf{onesol11}--\ref{onesol12}.
Now, considering the ratios of tau-functions appearing in the relations
\rf{phitau} and \rf{taucampos1}--\rf{taucampos2} one observes that in the
limit
$\rme^{\G_3(z_2)} \ra \infty$, such solution becomes the one-soliton
solution \rf{onesol11}--\rf{onesol12} again, but with the replacement
\be
\rme^{\G_1(z_1)} \ra \( \frac{z_1-z_2}{z_1+z_2} \) \, \rme^{\G_1(z_1)}
\ee
and with the ratios $\ttwo /\tone$, $\tthree /\tone$, $\tseven /\ttwo$,
$\tten / \tthree$, and $\televen /\tone$  changing signs. If we track the
soliton $\a_3$ instead of $\a_1$, we get the same results (interchanging
$\G_1 \leftrightarrow \G_3$), but the ratios of tau-functions changing
signs are $\ttwo /\tone$, $\tsix / \tone$, $\ttwenty /\ttwo$, and
$\ttwentyone /\tthree$. Therefore, comparing with \rf{shiftg1a}, we observe
that the lateral displacement and time delay for this two-soliton solution
is given by \rf{deltax1} and \rf{deltat}, respectively, with $n = 1$ and
$\omega = 1$. Consequently, the solitons of species $\a_1$ and $\a_3$
experience an attractive force.

\subsection{\mathversion{bold}Two solitons of species $\a_2$/$\a_3$}

Here we give the two-soliton solutions obtained by taking the group element
$\Sigma = \Sigma _{23} = \exp \left[ \rme^{\G_2(z_1)} V_{\a_2}
(a_{\a_2}^{\pm}, z_1) \right] \, \exp \left[ \rme^{\G_3(z_2)} V_{\a_3}
(a_{\a_3}^{\pm}, z_2) \right]$. The non-vanishing tau-functions are
\br
\tone &=& 1 - \frac{a^-_{\a_2} \, a^+_{\a_2}}{4} \, \rme^{2 \,
\Gamma_2(z_1)} - \frac{a^-_{\a_3} \, a^+_{\a_3}}{4} \, \rme^{2 \,
\Gamma_3(z_2)} \nonu \\*
&& \hspace{6em} + \frac{a^-_{\a_2} \, a^-_{\a_3} \, a^+_{\a_2} \,
a^+_{\a_3} \, (z_1 - z_2)^2}{16 \, (z_1 + z_2)^2} \, \rme^{2 \,
\Gamma_2(z_1) + 2 \, \Gamma_3(z_2)}, \\
\ttwo &=& 1 - \frac{a^-_{\a_2} \, a^+_{\a_2}}{4} \, \rme^{2 \,
\Gamma_2(z_1)} + \frac{a^-_{\a_3} \, a^+_{\a_3}}{4} \, \rme^{2 \,
\Gamma_3(z_2)} \nonu \\*
&& \hspace{6em} - \frac{a^-_{\a_2} \, a^-_{\a_3} \, a^+_{\a_2} \,
a^+_{\a_3} \, (z_1 - z_2)^2}{16 \, (z_1 + z_2)^2} \, \rme^{2 \,
\Gamma_2(z_1) + 2 \, \Gamma_3(z_2)}, \\
\tthree &=& 1 + \frac{a^-_{\a_2} \, a^+_{\a_2}}{4} \, \rme^{2 \,
\Gamma_2(z_1)} +  \frac{a^-_{\a_3} \, a^+_{\a_3}}{4} \, \rme^{2 \,
\Gamma_3(z_2)} \nonu \\*
&& \hspace{6em} + \frac{a^-_{\a_2} \, a^-_{\a_3} \, a^+_{\a_2} \,
a^+_{\a_3} \, (z_1 - z_2)^2}{16 \, (z_1 + z_2)^2} \, \rme^{2 \,
\Gamma_2(z_1) + 2 \, \Gamma_3(z_2)}, \\
\tfour &=& \frac{a^-_{\a_3} \, a^+_{\a_2} \, z_1}{z_1 + z_2} \,
\rme^{\Gamma_2(z_1) + \Gamma_3(z_2)}, \\
\tfive &=& a^-_{\a_2} \, \rme^{\Gamma_2(z_1)} + \frac{a^-_{\a_2} \,
a^-_{\a_3} \, a^+_{\a_3} \, (z_1 - z_2)}{4 \, (z_1 + z_2)} \,
\rme^{\Gamma_2(z_1) + 2 \, \Gamma_3(z_2)}, \\
\tsix &=& a^+_{\a_3} \, z_2 \, \rme^{\Gamma_3(z_2)} + \frac{a^-_{\a_2} \,
a^+_{\a_2} \, a^+_{\a_3} \, (z_1 - z_2) \, z_2}{4 \, (z_1 + z_2)} \,
\rme^{2 \, \Gamma_2(z_1) + \Gamma_3(z_2)}, \\
\tseven &=& \frac{a^-_{\a_2} \, a^+_{\a_3} \,z_2}{z_1 + z_2} \,
\rme^{\Gamma_2(z_1) + \Gamma_3(z_2)}, \\
\teight &=& a^+_{\a_2} \, \rme^{\Gamma_2(z_1)} - \frac{a^-_{\a_3} \,
a^+_{\a_2} \, a^+_{\a_3} \, (z_1 - z_2)}{4 \, (z_1 + z_2)} \,
\rme^{\Gamma_2(z_1) + 2 \, \Gamma_3(z_2)}, \\
\tnine &=& -\frac{a^-_{\a_3}}{z_2} \, \rme^{\Gamma_3(z_2)} +
\frac{a^-_{\a_2} \, a^-_{\a_3} \, a^+_{\a_2} \, (z_1 - z_2)}{4 \, z_2 \,
(z_1 + z_2)} \, \rme^{2\,\Gamma_2(z_1) + \Gamma_3(z_2)}, \\
\tten &=& -\frac{a^-_{\a_3} \, a^+_{\a_2} \, z_1}{z_2 \, (z_1 + z_2)} \,
\rme^{\Gamma_2(z_1) + \Gamma_3(z_2)}, \\
\televen &=& \frac{a^-_{\a_3} \, a^+_{\a_2}}{z_1 + z_2} \,
\rme^{\Gamma_2(z_1) + \Gamma_3(z_2)}, \\
\ttwelve &=& -\frac{a^-_{\a_2}}{z_1} \, \rme^{\Gamma_2(z_1)} +
\frac{a^-_{\a_2} \, a^-_{\a_3} \, a^+_{\a_3} \, (z_1 - z_2)}{4 \, z_1 \,
(z_1 + z_2)} \, \rme^{\Gamma_2(z_1) + 2 \, \Gamma_3(z_2)}, \\
\tthirteen &=& - \frac{a^-_{\a_2}}{z_1} \, \rme^{\Gamma_2(z_1)} -
\frac{a^-_{\a_2} \, a^-_{\a_3} \, a^+_{\a_3} \, (z_1 - z_2)}{4 \, z_1 \,
(z_1 + z_2)} \, \rme^{\Gamma_2(z_1) + 2 \, \Gamma_3(z_2)}, \\
\tfourteen &=& - a^+_{\a_3} \, \rme^{\Gamma_3(z_2)} + \frac{a^-_{\a_2} \,
a^{+}_{\a_2} \, a^+_{\a_3} \, (z_1 - z_2)}{4 \, (z_1 + z_2)} \, \rme^{2 \,
\Gamma_2(z_1) + \Gamma_3(z_2)}, \\
\tfifteen &=& - a^+_{\a_3} \, \rme^{\Gamma_3(z_2)} - \frac{a^-_{\a_2} \,
a^+_{\a_2} \, a^+_{\a_3} \, (z_1 - z_2)}{4 \, (z_1 + z_2)} \, \rme^{2 \,
\Gamma_2(z_1) + \Gamma_3(z_2)}, \\
\tsixteen &=& -\frac{a^-_{\a_2} \, a^+_{\a_3} \, z_1 \,z_2}{z_1 + z_2} \,
\rme^{\Gamma_2(z_1) + \Gamma_3(z_2)}, \\
\tseventeen &=& \frac{a^-_{\a_2} \, a^+_{\a_3} \, z_2^2}{z_1 + z_2} \,
\rme^{\Gamma_2(z_1) + \Gamma_3(z_2)}, \\
\teighteen &=& a^+_{\a_2} \, z_1 \, \rme^{\Gamma_2(z_1)} - \frac{a^-_{\a_3}
\, a^+_{\a_2} \, a^+_{\a_3} \, z_1 \, (z_1 - z_2)}{4 \, (z_1 + z_2)} \,
\rme^{\Gamma_2(z_1) + 2 \, \Gamma_3(z_2)}, \\
\tnineteen &=& a^+_{\a_2} \, z_1 \, \rme^{\Gamma_2(z_1)} + \frac{a^-_{\a_3}
\, a^+_{\a_2} \, a^+_{\a_3} \, z_1 \, (z_1 - z_2)}{4 \, ( z_1 + z_2)} \,
\rme^{\Gamma_2(z_1) + 2 \, \Gamma_3(z_2)}, \\
\ttwenty &=& a^-_{\a_3} \, \rme^{\Gamma_3(z_2)} + \frac{a^-_{\a_2} \,
a^-_{\a_3} \, a^+_{\a_2} \, (z_1 - z_2)}{4 \, (z_1 + z_2)} \, \rme^{2 \,
\Gamma_2(z_1) + \Gamma_3(z_2)}, \\
\ttwentyone &=& a^-_{\a_3} \, \rme^{\Gamma_3(z_2)} - \frac{a^-_{\a_2} \,
a^-_{\a_3} \, a^+_{\a_2} \, (z_1 - z_2)}{4 \, (z_1 + z_2)} \, \rme^{2 \,
\Gamma_2(z_1) + \Gamma_3(z_2)}.
\er
One can check that by taking the limit $\rme^{\G_3(z_2)} \ra 0$, such
solution become the one-soliton solution \rf{onesol21}--\rf{onesol22}. Now,
considering the ratios of tau-functions appearing in the relations
\rf{phitau} and \rf{taucampos1}--\rf{taucampos2} one observes that in the
limit
$\rme^{\G_3(z_2)} \ra \infty$, such solution becomes the one-soliton
solution \rf{onesol21}--\rf{onesol22} again, but with the replacement
\be
\rme^{\G_2(z_1)} \ra \( \frac{z_1-z_2}{z_1+z_2}\) \, \rme^{\G_2(z_1)}
\ee
and with the ratios $\ttwo /\tone$, $\tthree /\tone$, $\teight /\tthree$,
$\ttwelve / \ttwo$, and $\tthirteen /\tone$. If we track the soliton $\a_3$
instead of $\a_2$, we get the same results (interchanging $\G_2
\leftrightarrow \G_3$), but the ratios of tau-functions changing signs are
$\tthree /\tone$, $\tsix /\tone$, $\ttwenty / \ttwo$, and $\ttwentyone
/\tthree$. Therefore, comparing with \rf{shiftg1a}, we observe that the
lateral displacement and time delay for this two-soliton solution is given
by \rf{deltax1} and \rf{deltat}, respectively, with $n = 1$ and $\omega =
1$. Consequently, the solitons of species $\a_2$ and $\a_3$ experience an
attractive force.

\subsection{Summary of time delay results}

The calculations of the time delays of the two-soliton solutions have shown
that the solitons of the same species experience an attractive force. The
pair of solitons of species $\a_1$ and $\a_3$, as well as $\a_2$ and $\a_3$
also experience attractive forces. However, the pair of solitons $\a_1$ and
$\a_2$ suffer a repulsive force. In addition,  the force between solitons
of the same species is twice as strong than that between solitons of
different species. Consequently, the sign $\omega$ and the integer $n$
appearing in the time delay \rf{deltat} is determined by the scalar product
of the roots by the formula $(\a_i | \a_j) = \omega \, n$, by normalizing
the roots as $\a_i^2=2$.    Therefore, the time delay suffered by a soliton
of species $\a_i$, with rapidity $\theta_1$ when scattering with a soliton
of species $\a_j$, and rapidity $\theta_2$, with $\theta_1 >\theta_2$, is
given by the formula
\be
\Delta^{(i,j)}_1 (t) =  \frac{(\a_i | \a_j)}{m_i \cosh \theta_1}\,
\ln \left[ \tanh\( \frac{\theta_1-\theta_2}{2}\) \right].
\lab{deltax1conclu}
\ee

\vspace{1 cm} 

\noindent{\large{\bf Acknowledgements}}

\vspace{.2 cm}

We are grateful to H. Aratyn, J. F. Gomes, J. S\'anchez Guill\'en and
A. H. Zimerman for many helpful discussions. 
One of the authors (A.V.R) wishes to acknowledge the
warm hospitality of the Instituto de F\'\i sica Te\'orica -- IFT/UNESP,
S\~ao Paulo, Brazil, where his work on this paper was started, and the
financial support from FAPESP during his stay there in February-July 2000.
The research program of A.V.R. was supported in part by the Russian
Foundation for Basic Research under grant \#01--01--00201.  L.A.F is partially
supported by CNPq, and A.G.B is supported by FAPESP.

\newpage

\appendix

\sect{\mathversion{bold}The affine Kac-Moody algebra
$\widehat{\mathfrak{sl}_3(\mathbb C)}$}
\label{sec:app}

In this appendix we give the necessary information about the affine
Kac--Moody algebra $\widehat{\mathfrak{sl}_3(\mathbb C)}$. More details can
be found in the book by Kac \ct{kac} which we follow in our presentation
of basic facts and definitions, or in the review paper by Goddard and Olive
\ct{goddoliv} intended for physicists.

Recall that the Lie algebra $\mathfrak{sl}_3(\mathbb C)$ is the Lie algebra
formed by all $3 \times 3$ complex matrices with zero trace.
We use the standard choice for a Cartan subalgebra $\mathfrak h$ and for a
basis of the corresponding root system $\Delta$. Denote the three positive
roots by $\alpha_i$, $i = 1, 2, 3$, with $\alpha_a$, $a = 1, 2$, being the
simple roots and $\alpha_3 = \alpha_1 + \alpha_2$. The Cartan generators
are
\begin{equation}
H_1 = \left(\begin{array}{crc}
1 &  0 & 0 \\
0 & -1 & 0 \\
0 & 0 & 0
\end{array} \right), \qquad
H_2 = \left(\begin{array}{ccr}
0 & 0 & 0 \\
0 & 1 & 0 \\
0 & 0 & -1
\end{array} \right),
\end{equation}
and the basis vectors of the root subspaces corresponding to the positive
roots
are chosen as
\begin{equation}
E_{+\alpha_1} = \left(\begin{array}{ccc}
0 & 1 & 0 \\
0 & 0 & 0 \\
0 & 0 & 0
\end{array} \right), \qquad
E_{+\alpha_2} = \left(\begin{array}{ccc}
0 & 0 & 0 \\
0 & 0 & 1 \\
0 & 0 & 0
\end{array} \right), \qquad
E_{+\alpha_3} = \left(\begin{array}{ccc}
0 & 0 & 1 \\
0 & 0 & 0 \\
0 & 0 & 0
\end{array} \right).
\end{equation}
For negative roots we have
\begin{equation}
E_{-\alpha} = (E_{+\alpha})^T. \label{a.2}
\end{equation}
We use the invariant bilinear form on $\mathfrak{sl}_3(\mathbb C)$ defined
by
\begin{equation}
(x \, | \, y) = \mathrm{tr}(xy), \qquad x, y \in \mathfrak{sl}_3(\mathbb
C).
\end{equation}
The restriction of this form to the Cartan subalgebra $\mathfrak h$ is
nondegenerate, therefore it induces a nondegenerate bilinear form on
$\mathfrak h^*$ which we denote by $(\cdot \, | \, \cdot)$ as well. Note
that with this definition one has
\begin{equation}
(\alpha_1 | \alpha_1) = 2, \qquad (\alpha_2 | \alpha_2) = 2, \qquad
(\alpha_1 | \alpha_2) = -1.
\end{equation}

The affine Kac-Moody algebra $\widehat{\mathfrak{sl}_3(\mathbb C)}$ is
constructed in the following way. Consider the loop algebra
\begin{equation}
\mathcal L(\mathfrak{sl}_3(\mathbb C)) = \mathbb C[\zeta, \zeta^{-1}]
\otimes
\mathfrak{sl}_3(\mathbb C),
\end{equation}
where $\mathbb C[\zeta, \zeta^{-1}]$ is the algebra of Laurent polynomials
in $\zeta$.
An element of $\mathcal L(\mathfrak{sl}_3(\mathbb C))$ is a finite linear
combination of the elements of the form $\zeta^m \otimes x$, where $m \in
\mathbb Z$ and $x \in \mathfrak{sl}_3(\mathbb C)$. The structure of a Lie
algebra in $\mathcal L(\mathfrak{sl}_3(\mathbb C))$ is introduced by the
relation
\begin{equation}
\sbr{\zeta^m \otimes x}{\zeta^n \otimes y} = \zeta^{m+n} \otimes
\sbr{x}{y}.
\end{equation}
We identify the Lie algebra $\mathfrak{sl}_3(\mathbb C)$ with the
subalgebra of $\mathcal L(\mathfrak{sl}_3(\mathbb C))$ formed by the
elements of the form $1 \otimes x$, $x \in  \mathfrak{sl}_3(\mathbb C)$.
This allows us to write $\zeta^m x$ instead of $\zeta^m \otimes x$. Similar
identifications are used below.

Define a $\mathbb C$-valued 2-cocycle on $\mathcal
L(\mathfrak{sl}_3(\mathbb C))$ by
\begin{equation}
\psi(\zeta^m x, \, \zeta^n y) = m \, (x \, | \, y) \, \delta_{m+n,0}
\end{equation}
and denote by $\tilde \mathcal L(\mathfrak{sl}_3(\mathbb C))$ the extension
of $\mathcal L(\mathfrak{sl}_3(\mathbb C))$ by a one-dimensional center
associated to the cocycle $\psi$. Denote by $C$ the corresponding center
element, then the commutation relations in the Lie algebra $\tilde \mathcal
L(\mathfrak{sl}_3(\mathbb C)) = \mathcal L(\mathfrak{sl}_3(\mathbb C))
\oplus \mathbb C \, C$ are given by
\begin{equation}
\sbr{\zeta^m x}{\zeta^n y} = \zeta^{m+n} \sbr{x}{y} + m \, C \,
\delta_{m+n,
0}.
\label{a.1}
\end{equation}
Now denote by $\widehat{\mathfrak{sl}_3(\mathbb C)}$ the Lie algebra which
is obtained by adjoining to $\tilde \mathcal L(\mathfrak{sl}_3(\mathbb C))$
a derivation $D = \zeta (\mathrm d/\mathrm d \zeta)$. The commutation
relations for
the Lie algebra $\widehat{\mathfrak{sl}_3(\mathbb C)}$ are defined by
relations (\ref{a.1}) and by the equalities
\begin{equation}
\sbr{D}{\zeta^m x} = m \, \zeta^m x, \qquad \sbr{D}{C} = 0.
\end{equation}

The elements
\begin{equation}
H_1^m = \zeta^m H_1, \qquad H_2^m = \zeta^m H_2, \qquad E_\alpha^m =
\zeta^m
E_\alpha,
\end{equation}
where $m \in \mathbb Z$ and $\alpha \in \Delta$, together with the
elements $C$ and $D$ form a basis for $\widehat{\mathfrak{sl}_3(\mathbb
C)}$. These elements satisfy the commutation relations
\br
 \sbr{H_a^m}{H_b^n} &=& m \, (\a_a | \a_b) \, C \, \d_{m+n,0}, \label{a.3}
\\
 \sbr{H_a^m}{E_{\pm\a_i}^n} &=& \pm \, (\a_a | \a_i) \, E_{\pm\a_i}^{m+n},
\\
 \sbr{E_{\a_a}^m}{E_{-\a_a}^n} &=&  H_a^{m+n} + m \, C \,  \d_{m+n,0},
\\
 \sbr{E_{\a_3}^m}{E_{-\a_3}^n} &=& H_1^{m+n} + H_2^{m+n} + m\, C\,
\d_{m+n,0},
\\
 \sbr{E_{\a_1}^m}{E_{\a_2}^n} &=& E_{\a_3}^{m+n}, \\
 \sbr{E_{\a_3}^m}{E_{-\a_1}^n} &=& -\, E_{\a_2}^{m+n}, \\
\sbr{E_{\a_3}^m}{E_{-\a_2}^n} &=& E_{\a_1}^{m+n},
\lab{kmcomrel}
\er
where $a,b = 1,2$, $i = 1,2,3$ and $m, n \in \mathbb Z$.
The remaining non-vanishing commutation relations are obtained by using
relation (\ref{a.2}) which implies that
\begin{equation}
\sbr{E_{\a}^m}{E_{\b}^n} = - \sbr{E_{-\a}^m}{E_{-\b}^n}.
\end{equation}

We use in the paper the principal $\mathbb Z$-gradation of
$\widehat{\mathfrak{sl}_3(\mathbb C)}$ which can be defined by the grading
operator
\be
Q_{\mathrm{ppal}} = H_1 + H_2 + 3 D
\lab{gradop}
\ee
in the following way. Consider the following subspaces of
$\widehat{\mathfrak{sl}_3(\mathbb C)}$:
\begin{equation}
\mathfrak g_m = \{x \in \widehat{\mathfrak{sl}_3(\mathbb C)} \mid
[Q_{\mathrm{ppal}},\, x]
= m \, x \}, \qquad m \in \mathbb Z.
\end{equation}
It is easy to get convinced that
\begin{equation}
\widehat{\mathfrak{sl}_3(\mathbb C)} = \bigoplus_{m \in \mathbb Z}
\mathfrak g_m, \lab{ppalgrad}
\end{equation}
and, moreover,
\begin{equation}
\sbr{\gg_m}{\gg_n} \subset \gg_{m+n}.
\end{equation}
Hence, we really have a $\mathbb Z$-gradation of
$\widehat{\mathfrak{sl}_3(\mathbb C)}$. The subspace $\mathfrak g_0$ is a
actually a subalgebra of $\widehat{\mathfrak{sl}_3(\mathbb C)}$ given by
\be
\cgh_0 = \mathbb C \, H_1 \oplus \mathbb C \, H_2 \oplus \mathbb C \, C
\oplus \mathbb C \, D = \mathbb C \, H_1 \oplus \mathbb C \, H_2 \oplus
\mathbb C \, C  \oplus \mathbb C \, Q_{\mathrm{ppal}},
\lab{zgsa}
\ee
and for other subspaces $\mathfrak g_m$ we have
\begin{eqnarray}
\cgh_{3m} &=& \mathbb C \, H_1^m \oplus \mathbb C \, H_2^m, \quad m \ne 0,
\\
\cgh_{3m+1} &=& \mathbb C \, E_{\a_1}^m \oplus \mathbb C \, E_{\a_2}^m
\oplus \mathbb C \, E_{-\a_3}^{m+1}, \\
\cgh_{3m+2} &=& \mathbb C \, E_{-\a_1}^{m+1} \oplus \mathbb C \,
E_{-\a_2}^{m+1} \oplus \mathbb C \, E_{\a_3}^{m}. \lab{a.4}
\end{eqnarray}

Among the representations of $\cgh$ there are three which play an important
role. They are the fundamental highest weight representations. Each one has
a highest weight state $| \, \l_j \, \rangle$, $j = 0, 1, 2$, satisfying
\be
H_a \, | \, \l_0 \, \rangle = 0, \qquad H_a \, | \, \l_b \, \rangle =
\d_{a,b} | \, \l_b \, \rangle, \qquad C \, | \, \l_j \, \rangle = | \, \l_j
\, \rangle \lab{fundrep1}
\ee
for $a,b = 1,2$ and $j = 0,1,2$. Such states are annihilated by all
positive grade subspaces
\be
\cgh_m \, \ket{\l_j} = 0, \qquad \qquad m > 0,
\lab{fundrep2}
\ee
and all the states of the representation spaces are obtained by acting on
the corresponding highest weight state $\ket{\l_j}$, with negative grade
generators. The representation spaces can be supplied with a scalar product
with respect to which
\begin{eqnarray}
& (H^m_a)^\dagger = H^{-m}_a, \qquad (E^m_{\alpha})^\dagger =
E^{-m}_{-\alpha}, \\
& C^\dagger = C, \qquad D^\dagger = D.
\end{eqnarray}
It follows from \rf{zgsa}--\rf{a.4} that in this case
\be
(\mathfrak g_m)^\dagger = \mathfrak g_{-m},
\ee
and one has
\be
\bra{\l_j}\, \cgh_{-m}  = 0, \qquad \qquad m > 0.
\lab{fundrep3}
\ee

Recall that $\gg_0$ is a subalgebra of $\asl3$. It is convenient also to
consider two additional subalgebras
\be
\gg_{<0} = \bigoplus_{m > 0} \gg_m, \qquad \gg_{>0} = \bigoplus_{m > 0}
\gg_{-m}.
\ee
These subalgebras and the corresponding Lie groups play important role in
the dressing transformation method.

\newpage 

\sect{Mathematica program implementing Hirota method}
\label{sec:app2}

In this appendix we give the source of a {\sc Mathematica} program which
allow to obtain and test soliton solutions of equations
(\ref{eqmot1})--(\ref{eqmot2}) using the Hirota method.

First of all note that to use it one has fix the number of solitons and the
soliton species. This information is saved in the variable {\tt
n} and in the list {\tt s} respectively. The Hirota ansatz for
tau-functions is given by the function {\tt t}. The corresponding
coefficents are collected into the arrays {\tt b}. The function {\tt eqs}
extracts from the equations for the tau-functions the terms with the fixed
total power of $\rme^{\Gamma_i}$. The function {\tt calcbs} calculates the
coefficients of the Hirota ansatz iteratively. The function {\tt testsol}
test that the equations for the tau-functions are satified by the solution
found. The concrete code follows.
\small
\begin{verbatim}
g[i_, z_, xp_, xm_] := m[i] (z xp - xm/z)/2;
m[3] = m[1] + m[2];
c[l__][i_] := 1 /; i <= 3 && Plus[l] == 0;
c[l__][i_] := 0 /; i > 3 && Plus[l] == 0;
Ind[p_] := Select[AllInd, Plus @@ #1 == p &];
b[l__] := Array[c[l], 21];
t[k_, xp_, xm_] :=
  Sum[Product[(e[i]Exp[g[s[[i]], z[i], xp, xm]])^AllInd[[j]][[i]],
    {i, n}](c @@ AllInd[[j]])[k], {j, Length[AllInd]}];
eqs[p_, l_] :=
    Block[{temp = Ind[p][[l]]},
      Table[D @@
            Prepend[Table[{e[i], temp[[i]]}, {i, n}],
              taueq[k, xp, xm]] /. {e[i_] -> 0}, {k, 21}]];
calcbs :=
    Block[{temp, outf},
      Off[Solve::svars];
      AllInd = Flatten[
        Table @@ Join[{Table[i[k], {k, n}]}, Table[{i[k], 0, 2}, {k, n}]],
          n - 1];
      Do[
        temp = b @@ Ind[p][[k]];
        Evaluate[b @@ Ind[p][[k]]] =
          Factor[temp /. Solve[eqs[p, k] == 0, temp][[1]]], {p, 2 n},
          {k, Length[Ind[p]]}];
      outf = ToString[n] <> "sol" <> ToString[SequenceForm @@ s] <> ".m";
      If[Length[FileNames[outf]] != 0, DeleteFile[outf], Null];
      Save[outf, c];
      Print["The results of calculations were saved in file " <> outf];
      On[Solve::svars]];
testsol :=
    If[Table[Simplify[taueq[k, xp, xm] /. {xm -> 0, xp -> 0}], {k, 21}]
        == Table[0, {21}],
      Print["Equations are satisfied"],
      Print["Equations are not satisfied"]];
\end{verbatim}
\normalsize
Below we give the script of a sample {\sc Mathematica} session. It is
supposed that the above code is in the file {\tt gensol.m} and the
equations for the tau-functions are in the file {\tt taueqs.m}.
\begin{verbatim}
In[1]:= << taueqs.m

In[2]:= << gensol.m

In[3]:= n = 2;

In[4]:= s = {1, 3};

In[5]:= calcbs
The results of calculations were saved in file 2sol13.m

In[6]:= testsol
Equations are satisfied
\end{verbatim}
In conclusion note that the program above works in principle for any number
of solitons, but already for three-soliton solutions the procedure of
testing the solution is very time consuming.

\end{document}